\documentstyle[12pt]{article}

\def\Let@{\relax\iffalse{\fi\let\\=\cr\iffalse}\fi}
\def\vspace@{\def\vspace##1{\crcr\noalign{\vskip##1\relax}}}
\def\multilimits@{\bgroup \vspace@\Let@
 \lineskip 2pt
 \lineskiplimit\lineskip
 \vbox\bgroup\ialign\bgroup\hfil$\scriptstyle{##}$\hfil\crcr}
\def\Sb{_\multilimits@}
\def\endSb{\crcr\egroup\egroup\egroup}
\def\Sp{^\multilimits@}

\def\text{\rm }

\begin{document}

\pagestyle{empty} \noindent {\small USC-99/HEP-B4\hfill \hfill
hep-th/9907087 }\newline
{\small \hfill }

{\vskip 0.7cm}

\begin{center}
{\Large {\bf String Theory on $AdS_3\,\,$Revisited$^{*}$}} \\[0pt]

{\vskip 0.5cm}

{\bf I. Bars$^{\dag },$ C. Deliduman$^{\ddag }$ and D. Minic}$^{\natural }$%
{\bf \ } {\Large \ \\[0pt]
}

{\vskip 0.4cm}

Department of Physics and Astronomy

University of Southern California

Los Angeles, CA 90089-0484

{\vskip 0.5cm}

{\bf ABSTRACT}
\end{center}

We discuss string theory on $AdS_3\times S^3\times M^4$ with particular
emphasis on unitarity and state-operator correspondence. The AdS-CFT
correspondence, in the Minkowski signature, is re-examined by taking into
account the only allowed unitary representation: the principal series module
of the affine current algebra SL(2,R) supplemented with zero modes. Zero
modes play an important role in the description of on-shell states as well
as of windings in space-time at the AdS$_3$ boundary. The theory is
presented as part of the supersymmetric WZW model that includes the
supergroup SU$\left( 2/1,1\right) $ (or OSp$\left( 4/2\right) $ or D$\left(
2,1;\alpha \right) $) with central extension $k$. A free field
representation is given and the vertex operators are constructed in terms of
free fields in SL(2,R) principal series representation bases that are
labeled by position space or momentum space at the boundary of AdS$_3$. The
vertex operators have the correct operator products with the currents and
stress tensor, all of which are constructed from free fields, including the
subtle zero modes. It is shown that as $k\rightarrow \infty $, AdS$_3$ tends
to flat 3D-Minkowski space and the AdS$_3$ vertex operators in momentum
space tend to the vertex operators of flat 3D-string theory (furthermore the
theory readjusts smoothly in the rest of the dimensions in this limit).

\vfill
\hrule width 6.7cm \vskip 2mm

{$^{*}$ {\small Research partially supported by the US. Department of Energy
under grant number DE-FG03-84ER40168 and by the National Science Foundation
under grant number NSF9724831 for collaborative research between USC and
Japan}}.

{$^{\dag }$ {\small bars@physics.usc.edu, }$^{\ddag }$ {\small %
cdelidum@physics.usc.edu, }}$^{\natural }$ {\small minic@physics.usc.edu}

\vfill\eject

\setcounter{page}1\pagestyle{plain}

\section{Introduction}

There has been much discussion on the topic of a string propagating on AdS$%
_{3}$ curved spacetime \cite{BRFW}-\cite{berkvw}. The early interest was due
to the challenge of solving exactly the problem of a string theory in curved
spacetime {\it with Minkowski signature}. At the early stages this effort
revealed some problems with ghosts, despite the Virasoro constraints. The
solution to the problem was presented several years later in \cite{Bars} 
\cite{Str95} where it was shown that there were two simple but essential
points that were missed in earlier investigations: (i) the correct unitary
representation and (ii) zero modes. Both of these are provided naturally by
the explicit analysis of the non-compact SL$\left( 2,R\right) $ WZW model
but it was easy to miss them in the earlier abstract current algebra
approach that included wrong assumptions by analogy to the compact SU$\left(
2\right) $. While each point is an independent feature of the model they are
both needed to correctly describe the string in AdS$_{3}$ space.

The first point is that the WZW model based on SL(2,R), in the absence of
certain zero modes (in analogy to SU$\left( 2\right) $), permits only the
unitary representation called the principal series for which $j(j+1)\leq
-1/4 $ (or $j=-1/2+is$ where $s$ is real). This is similar to saying that
the model ${\bf L=r\times p}$ for angular momentum permits only integer
quanta for angular momentum; half integer quanta cannot occur in this model.
Similarly, the discrete series or the supplementary series modules of affine
SL$\left( 2,R\right) $ current algebra cannot occur in the WZW model. The
ghost problems do not arise in the principal series module, they only arise
in the discrete series module which was wrongly assumed to be part of the
model when using abstract current algebra in the investigations prior to
1995. In recent investigations \cite{perry} - \cite{kutseib} the discrete
series module resurfaced in connection with the AdS-CFT conjecture, however
this is unsatisfactory since the relation to the underlying string theory
remains obscure and the corresponding vertex operators lack unitarity or
state-operator correspondence. We show that there is a subtle solution that
incorporates unitarity and state-operator correspondence, consistent with
the underlying unitary string theory spectrum given in \cite{Bars} \cite
{Str95}.

The second point is that a lightcone-type momentum zero mode $p_L^{-}$ (and
a similar $p_R^{+}$) was also missed in the old abstract current algebra
approach. The zero mode is needed to satisfy the Virasoro mass shell
condition ($L_0=a$) for left movers in the form 
\begin{equation}
L_0=p_L^{+}p_L^{-}-j(j+1)/(k-2)+{\it {integer}=a\leq 1}  \label{L0}
\end{equation}
(and a similar one for right movers). In the absence of $p_L^{-}$ the mass
shell condition cannot be satisfied with the principal series ($%
-j(j+1)=\frac 14+s^2$) when the positive integer is non-zero (string
excitation level) . The zero mode term $p_L^{+}p_L^{-}$ provides the only
negative contribution ( $p_L^{+}p_L^{-}=-p_{L0}^2+p_{L1}^2<0$) thereby
making it possible to satisfy the on shell condition in the same manner of a
string in flat spacetime (in fact an excited string in flat spacetime also
cannot be put on shell if the $p^{-}$ zero mode is absent: $%
p^{+}p^{-}+(p_2)^2+{\it {integer}=1}$). However, the presence of the $%
p_L^{-} $ zero mode in curved spacetime is non-trivial and it requires a
quantization condition due to the periodicity of a closed string $x\left(
\tau ,\sigma +2\pi \right) =x\left( \tau ,\sigma \right) $. It was shown in 
\cite{Bars} \cite{Str95} that due to this monodromy of the SL$\left(
2,R\right) $ currents, unless $p_L^{-}$ vanishes, one must have a negative
integer for the combination $p^{+}p^{-}$, that is 
\begin{equation}
p_L^{+}p_L^{-}=p_R^{+}p_R^{-}=-r,\quad r\in Z_{+}\,\,.
\end{equation}
For the compact SU$\left( 2\right) $ (as opposed to the non-compact SL$%
\left( 2,R\right) $) the zero modes $p_L^{-},p_R^{+}$ are not needed, and
are taken to vanish. This is why they were missed in the early
investigations of SL$\left( 2,R\right) $. With these conditions the complete
on-shell spectrum of the theory was computed, the no-ghost theorem was
proven (after the Virasoro constraints) and the exact unitary solution of
the SL$\left( 2,R\right) $ WZW model was established \cite{Bars} \cite{Str95}%
.

The recent interest in the topic is due to the AdS-CFT conjecture \cite
{Maldacena}. The AdS$_3$ string provides one of the rare exact conformal
field theories in which one could possibly verify the conjecture for a full
superstring theory as opposed to the low energy supergravity limit. It is
important to take into account the unitarity of the model as a string on AdS$%
_3$ if this correspondence is to be a meaningful one. Recent papers \cite
{GKS} - \cite{kutseib} made progress in providing an AdS$_3$-CFT map by
suggesting and investigating certain vertex operators constructed in AdS$_3$
string theory that correspond to operators in the boundary CFT theory. One
interesting overlap with the previous work \cite{Bars} \cite{Str95} involves
the quantized zero modes $p_L^{-},p_R^{+}$ described above: they are related
to the winding numbers on the AdS boundary $\oint d\gamma /\gamma $ , $\oint
d\bar{\gamma}/\bar{\gamma}$ in the language of \cite{GKS} and \cite{kutseib}
(this will be described more fully below). However, the recent papers did
not incorporate the unitarity conditions given in \cite{Bars} \cite{Str95}
and the operator-physical state correspondence that is standard and
desirable in string theory is obviously lacking. For these reasons, while
the CFT-AdS ideas in recent papers are very tantalizing, the connection to
the underlying physical string theory remains to be established. We will
show that we are not far from this when we insist on unitarity and use only
the principal series of the affine current algebra supplemented with zero
modes.

In this paper we briefly review the construction of \cite{Bars} \cite{Str95}
and present it in the context of a superstring on $AdS_3\times S^3\times M^4$
partially described by the WZW model that includes the supergroup SU$\left(
2/1,1\right) $ (or OSp$\left( 4/2\right) $ or D$\left( 2,1;\alpha \right) $)
with central extension $k$. We emphasize the two points mentioned above: (i)
only the principal series of the affine SL$\left( 2,R\right) $ is allowed
and (ii) it must be supplemented by the quantized zero modes $%
p_L^{-},p_R^{+} $.

We give the free field representation of the model and then construct the
vertex operators in terms of free fields in two SL($2,R$) bases. The SL(2,R)
labels on these representations have the interpretation of either position
or momentum at the AdS boundary. The momentum space version was introduced
in \cite{Str95} and subsequently completed in an unpublished work by two of
us \cite{bardel}; its details will be presented here. The position space
version is related to the momentum space one by ordinary Fourier
transformation. It is also related to an analytic continuation of the vertex
operator for a string on $H_{+}^3=$SL$\left( 2,C\right) /SU\left( 2\right) $
used in the recent literature \cite{teschner} \cite{teschner2} (take $%
H_{+}^3\rightarrow $SL$\left( 2,R\right) $ after Euclidean to Minkowski
continuation), but only in the sector in which the zero modes $%
p_L^{-},p_R^{+}$ vanish.

We also go one step further by constructing the vertex operator from free
fields such that the AdS string coordinates $\phi \left( z,\bar{z}\right)
,\gamma \left( z,\bar{z}\right) ,\bar{\gamma}\left( z,\bar{z}\right) $ in
curved space are themselves constructed from left and right moving free
fields. We show that the free field representation of the vertex operators
satisfy the correct operator products with the SL$\left( 2,R\right) $
currents and the stress tensor, all constructed from left/right free fields.
Using the free field construction we show that the quantized zero modes $%
p_L^{-},p_R^{+}$ are proportional to the winding numbers $\oint d\gamma
/\gamma $ , $\oint d\bar{\gamma}/\bar{\gamma}$ at the boundary of AdS$_3$.

We show one other desirable property of the vertex operator, namely that it
becomes the flat string vertex operator exp$\left( ip_\mu X^\mu \left( z,%
\bar{z}\right) \right) $ when the AdS$_3$ space tends to flat space as the
central extension grows $k\rightarrow \infty $.

Using the new unitary vertex operators we re-examine the AdS-CFT
correspondence discussed in recent papers. We show that there is operator-
physical state correspondence. Also we show that all the main arguments of
the CFT-AdS correspondence can be reformulated by using only the unitary
vertex operators.

There remains to compute in strictly string theoretical language the various
correlation functions or operator products that would verify or refine the
AdS-CFT correspondence. In principle our free field vertex operators can be
used to compute any correlation function, but we leave these computations to
future papers.

\section{SU$\left( 2/1,1\right) $ WZW Model and free fields}

\subsection{Comments on the supergroup approach}

The WZW model based on the {\it non-compact} supergroup SU$\left(
2/1,1\right) $ at level $k>0$ has one timelike and five spacelike bosonic
coordinates; it also has four timelike and four spacelike fermionic degrees
of freedom (see next paragraph). Therefore it has ghosts that come from
timelike bosonic and fermionic modes. An additional indication that it
contains ghosts is that, as for any SU$\left( N/N\right) $ current algebra 
\cite{ibfree}, its Virasoro central charge is $c=-2$ (see footnote 2 of the
first paper listed in \cite{ibfree}). Indeed as argued in \cite{berkvw} this
is a model for a superstring on AdS$_3\times S^3$ with Ramond-Ramond flux,
but its degrees of freedom include Faddeev-Popov ghosts that arise in the
quantization procedure. The physical sector of the model can be obtained by
applying an elaborate set of constraints, as discussed in \cite{berkvw}. A
simpler picture is to consider $N=4$ local superconformal symmetry. Such a
local symmetry provides one bosonic constraint (usual Virasoro) and four
fermionic constraints that match the number of timelike bosonic/fermionic
degrees of freedom, thus leading to a ghost-free physical sector. Assuming
the sufficiency of these constraints the WZW model for the supergroup SU$%
\left( 2/1,1\right) $ would correspond to a physical superstring theory
whose interpretation is obtained through the arguments of \cite{berkvw}.

The counting of timelike degrees of freedom and constraints (gauge
invariances) that remove ghosts is an essential first step for determining
if WZW models (or gauged WZW models) based on supergroups can be physical
superstring theories. A quick way to arrive at the counting is to consider
the signature of the fields in the Lagrangian or equivalently the signature
in the operator products of the currents $J^A$. The kinetic term in the
Lagrangian or the double pole in the operator product $\left( z-w\right)
^{-2}$ is proportional to the central extension times the supergroup Killing
metric $\frac k2$Str$\left( T^AT^B\right) $ where $T^A$ is a graded
supermatrix that represents the generators in the fundamental representation%
\footnote{%
Under hermitian conjugation the SU$\left( 2/1,1\right) $ matrices satisfy $%
\left( T^A\right) ^{\dagger }=CT^AC^{-1}$ with 
\begin{equation}
C=\left( 
\begin{array}{cc}
1_2 & 0 \\ 
0 & \tau _2
\end{array}
\right)
\end{equation}
where $1_2$ is the 2$\times 2$ identity matrix and $\tau _2$ is the Pauli
matrix (by taking $\tau _2$ we have chosen the SL$\left( 2,R\right) $ rather
than the SU$\left( 1,1\right) $ basis). Then the supermatrix $T^a$ with $%
a=1,2,3$ that correspond to SU$\left( 2\right) $ has non-zero entries $\tau
^a$ (Pauli matrices) in the upper left 2$\times 2$ block, the supermatrix $%
T^\mu $ with $\mu =0,1,2$ that correspond to the SU$\left( 1,1\right) $=SL$%
\left( 2,R\right) $ subalgebra has non-zero entries $\sigma ^\mu \equiv
\left( \tau _2,i\tau _1,-i\tau _3\right) $ in the lower right 2$\times 2$
block. The supermatrices $T^\alpha $ that correspond to the eight fermionic
parameters (4 complex fermions) have non-zero entries in the off diagonal
blocks.}. The timelike/spacelike signature of the bosonic/fermionic string
coordinate (or supergroup parameter) $X^A\left( z\right) $ that is
associated with $T^A$ is directly determined by the sign of $\frac k2$Str$%
\left( T^AT^B\right) $. For SU$\left( 2/1,1\right) $ with $k>0,$ the SU$%
\left( 2\right) $ subalgebra matrices give a positive signature $\frac k2$Str%
$\left( T^aT^b\right) =k\delta ^{ab}$ ; hence the SU$\left( 2\right) $ group
parameters correspond to spacelike string coordinates. For the SU$\left(
1,1\right) $ subalgebra one has $\frac k2$Str$\left( T^\mu T^\nu \right)
=-k\eta ^{\mu \nu }$ where the extra minus sign comes from the definition of
the supertrace, and the SL$\left( 2,R\right) $ Killing metric $\eta ^{\mu
\nu }$ is given by $\eta ^{00}=-\eta ^{22}=-\eta ^{11}=1$. Therefore, the SU$%
\left( 1,1\right) $ parameters correspond to two spacelike and one timelike
coordinates. Similarly, for the fermionic parameters one finds 4 spacelike
and 4 timelike string degrees of freedom.

\subsubsection{ Supergroups OSp$\left( 4/2\right) $ and D$\left( 2,1;\protect%
\alpha \right) $}

As another example consider the supergroup OSp$\left( 4/2\right) $ with
compact SO$\left( 4\right) $ and non-compact Sp$\left( 2\right) =SL\left(
2,R\right) $. In this case there are 8 spacelike and one timelike bosonic
coordinates and 4 spacelike and 4 timelike fermionic ones. The supergroup D$%
\left( 2,1;\alpha \right) $ also has the same counting (at $\alpha =1$ it
becomes OSp$\left( 4/2\right) $ ). Therefore, physical string theories can
be constructed with the supergroups SU$\left( 2/1,1\right) $, OSp$\left(
4/2\right) $ and D$\left( 2,1;\alpha \right) $ provided the remaining
spacelike degrees of freedom added and then $N=4$ superconformal constraints
are imposed. For other supergroups one may consider gauged WZW models based
on supergroups that give only one timelike coordinate (see \cite{ibonetime}
for the purely bosonic sector) supplemented by appropriate fermionic
constraints that follow from gauge invariances.

\subsection{Free fields and reduction to bosonic AdS$_3$}

Returning to SU$\left( 2/1,1\right) $, the group element can be parametrized
at the critical point in the usual form $G\left( z,\bar{z}\right) =G_L\left(
z\right) G_R\left( \bar{z}\right) $. Then each of the $G_{L,R}$ can most
generally be parametrized in the ``triangular'' form 
\begin{equation}
G_L\left( z\right) =\left( 
\begin{array}{cc}
1 & 0 \\ 
\theta _L & 1
\end{array}
\right) \left( 
\begin{array}{cc}
H_L & 0 \\ 
0 & g_L
\end{array}
\right) \left( 
\begin{array}{cc}
1 & \tilde{\theta}_L \\ 
0 & 1
\end{array}
\right)  \label{gl}
\end{equation}
where $\left( H_L\left( z\right) \right) _i^j$ and $\left( g_L\left(
z\right) \right) _\alpha ^\beta $ are group elements in the adjoint
representation $\left( 1,0\right) $ and $\left( 0,1\right) $ of the SU$%
\left( 2\right) \otimes SL\left( 2,R\right) $ subgroup, $\left( \theta
_L^{\,}\left( z\right) \right) _\alpha ^{\,i}$ is a 2$\times 2$ matrix of 4
complex fermions in the $\left( \frac 12,\frac 12\right) $ representation,
and $\tilde{\theta}_L\left( z\right) =\theta _L^{\dagger }\sigma _2$. In
this notation canonical conjugates are determined for the WZW model, and its
currents are parametrized in terms of them as in \cite{ibfree}\cite{berkvw} 
\begin{eqnarray}
SU\left( 2\right) &:&W^a=\theta _\alpha ^{\,i}\left( \frac{\tau ^a}2\right)
_i^j\pi _j^\alpha +J^a \\
SL\left( 2,R\right) &:&W^\mu =-\theta _\alpha ^{\,i}\left( \frac{\sigma ^\mu 
}2\right) _\beta ^\alpha \pi _i^\beta +J^\mu \\
coset &:&W_i^\alpha =\pi _i^\alpha \\
\bar{W}_\alpha ^i &=&-\theta _\alpha ^{\,j}\theta _\beta ^{\,i}\pi _j^\beta
+\theta _\beta ^{\,i}\left( \sigma _\mu \right) _\alpha ^\beta J^\mu
-J^a\left( \tau _a\right) _j^i\theta _\alpha ^{\,j}+k\partial _z\theta
_\alpha ^{\,j}
\end{eqnarray}
The fields $\left( \theta _L^{\,}\left( z\right) \right) _\alpha ^{\,i}$ and
its canonical conjugate $\left( \pi _L\left( z\right) \right) _i^\alpha $
are just free fields akin to Wakimoto's \cite{wakim} representation ($\gamma
\sim \theta _\alpha ^i$ and $\beta \sim \pi _i^\alpha \left( z\right) $),
but they are fermions as well as being matrices in the present case. Also $%
\left( \partial H_LH_L^{-1}\right) _i^j=\left( \tau _a\right) _i^jJ^a\left(
z\right) $ is an SU$\left( 2\right) $ current at level $k-2$ and $\left(
\partial g_Lg_L^{-1}\right) _\alpha ^\beta =\left( t_\mu \right) _\alpha
^\beta J^\mu \left( z\right) $ is an SL$\left( 2,R\right) $ current at level 
$-k-2$ (see \cite{ibfree} for an explanation of the shift in $k)$. The
Sugawara stress tensor is \cite{ibfree}\cite{berkvw} 
\begin{equation}
T_{++}\left( z\right) =:\pi _i^\alpha \partial \theta _\alpha ^{\,i}:+\frac
1k:\left( J^aJ_a-J^\mu J_\mu \right) :  \label{superT}
\end{equation}
This shows that the fermions $\theta ,\pi $ are free fields and that the
analysis of the model is reduced to solving the left/right factorized
bosonic WZW model based on SU$\left( 2\right) \times $SL$\left( 2,R\right) $
at levels $k-2$ and $-k-2$ respectively. The SU$\left( 2\right) $ part is
well known, therefore we concentrate on the purely bosonic SL$\left(
2,R\right) $ part and discuss the purely bosonic string on the AdS$_3$
background by itself. This involves the purely bosonic degrees of freedom $%
g_L\left( z\right) $ and the corresponding currents $J^\mu \left( z\right) $
(and their right moving counterparts $g_R\left( \bar{z}\right) ,\tilde{J}%
^\mu \left( \bar{z}\right) )$ to which we will return in the following
sections.

Note that the combined contribution of the bosonic currents $J^{a},J^{\mu }$
to the Virasoro central charge $c_{bosons}$ is independent of $k$%
\begin{eqnarray}
c_{SU\left( 2\right) } &=&\frac{3\left( k-2\right) }{\left( k-2\right) +2}=3-%
\frac{6}{k},\quad c_{SL\left( 2,R\right) }=\frac{3\left( -k-2\right) }{%
\left( -k-2\right) +2}=3+\frac{6}{k} \\
c_{bosons} &=&c_{SU\left( 2\right) }+c_{SL\left( 2,R\right) }=6,\quad
c_{fermions}=-8,\quad c_{total}=-2.
\end{eqnarray}
Therefore the value of the central extension $k$ can be changed arbitrarily
without changing the central charge $c_{bosons}$. This feature will allow us
to consider the limit $k\rightarrow \infty $ for which AdS$_{3}$ tends to
flat 3D Minkowski space while the rest of the theory adjusts to this limit
smoothly. In this limit we will show that our AdS$_{3}$ vertex operator
tends to the vertex operator in flat 3D space :$\exp \left( ik_{\mu }X^{\mu
}\left( z,\bar{z}\right) \right) :$ which is obviously a desired property
for the correct vertex operator.

The same procedure can be applied to OSp$\left( 4/2\right) $ and D$\left(
2,1;\alpha \right) $ for which again the fermions are free fields \cite
{ibfree} and the non-trivial part is the string on AdS$_{3}$. These
supergroups can be used to describe superstrings on AdS$_{3}\times
S^{3}\times S^{3}$. In this case the bosonic part corresponds to SL$\left(
2,R\right) \times $SU$\left( 2\right) _{1}\times $SU$\left( 2\right) _{2}$
at levels $-k-2$, $\left( 1+\alpha \right) k-2$ and $\left( 1+1/\alpha
\right) k-2$ respectively ($\alpha =1$ is for OSp$\left( 4/2\right) $). One
can then see that $c_{bosons}=9$, $c_{fermi}=-8$, $c_{tot}=1$ are again
independent of $k$ or $\alpha $ and therefore these parameters can be taken
to various limits to better understand the structure of the model.

\subsection{Bosonic string on AdS$_{3}$}

We now discuss the purely bosonic string in an AdS$_3$ background. This
problem was already solved satisfactorily in \cite{Bars} \cite{Str95} but
the lessons learned there remain to be incorporated in the study of the
AdS-CFT conjecture, which we will address in later sections. In this section
we outline the main important features that relate to unitarity and to zero
modes (the notation is slightly different here as compared to \cite{Bars} 
\cite{Str95})

To parametrize the group element in the $j=1/2$ representation we use the
matrix representation $t^{\mu }=\sigma ^{\mu }/2$ given in footnote 1. Using
the lightcone type combinations 
\begin{eqnarray}
t^{+} &=&t^{1}+t^{0}=i\frac{\tau ^{1}}{2}+\frac{\tau ^{2}}{2}=i\left( 
\begin{array}{cc}
0 & 0 \\ 
1 & 0
\end{array}
\right) \\
t^{-} &=&t^{1}-t^{0}=i\frac{\tau ^{1}}{2}-\frac{\tau ^{2}}{2}=i\left( 
\begin{array}{cc}
0 & 1 \\ 
0 & 0
\end{array}
\right) \\
t^{2} &=&-i\frac{\tau ^{3}}{2}=-i\left( 
\begin{array}{cc}
\frac{1}{2} & 0 \\ 
0 & -\frac{1}{2}
\end{array}
\right)
\end{eqnarray}
we can always parametrize the SL$\left( 2,R\right) $ group element in the
triangular form 
\begin{eqnarray}
g\left( z,\bar{z}\right) &=&e^{it^{+}\gamma ^{-}\left( z,\bar{z}\right)
}e^{it_{2}\phi \left( z,\bar{z}\right) }e^{it^{-}\gamma ^{+}\left( z,\bar{z}%
\right) } \\
&=&\left( 
\begin{array}{cc}
1 & 0 \\ 
-\gamma ^{-}\left( z,\bar{z}\right) & 1
\end{array}
\right) \left( 
\begin{array}{cc}
e^{\frac{1}{2}\phi \left( z,\bar{z}\right) } & 0 \\ 
0 & e^{-\frac{1}{2}\phi \left( z,\bar{z}\right) }
\end{array}
\right) \left( 
\begin{array}{cc}
1 & -\gamma ^{+}\left( z,\bar{z}\right) \\ 
0 & 1
\end{array}
\right)  \nonumber \\
&=&\left( 
\begin{array}{cc}
e^{\frac{1}{2}\phi } & -\gamma ^{+}e^{\frac{1}{2}\phi } \\ 
-\gamma ^{-}e^{\frac{1}{2}\phi } & e^{-\frac{1}{2}\phi }+e^{\frac{1}{2}\phi
}\gamma ^{+}\gamma ^{-}
\end{array}
\right)  \label{figamma}
\end{eqnarray}
Note that all the fields $\phi \left( z,\bar{z}\right) $, $\gamma ^{-}\left(
z,\bar{z}\right) $, $\gamma ^{+}\left( z,\bar{z}\right) $ are real. $\gamma
^{\pm }\left( z,\bar{z}\right) $ are lightcone combinations $\gamma ^{\pm
}=\gamma ^{1}\pm \gamma ^{0}$. The SL$\left( 2,R\right) $ WZW Lagrangian
takes the form 
\begin{equation}
L =\left( k+2\right) \left[ \partial \phi \bar{\partial}\phi + e^{\phi
}\partial \gamma ^{+}\bar{\partial}\gamma ^{-} \right]
\end{equation}
This shows that it describes a string with one time-like coordinate and two
spacelike coordinates propagating on the AdS$_{3}$ background. The boundary
of AdS$_{3}$ is defined by $\phi \rightarrow \infty $.

The classical solution for the fields $\phi \left( z,\bar{z}\right) $, $%
\gamma ^{-}\left( z,\bar{z}\right) $, $\gamma ^{+}\left( z,\bar{z}\right) $
are extracted from the general classical solution for the group element in
the form 
\begin{equation}
g\left( z,\bar{z}\right) =g_{L}\left( z\right) g_{R}\left( \bar{z}\right)
\label{LR}
\end{equation}
where $g_{L,R}$ are arbitrary SL$\left( 2,R\right) $ group elements. Note
that the $g_{L}\left( z\right) $ that appears here is identified with the $%
g_{L}\left( z\right) $ in the parametrization of the SU$\left( 2/1,1\right) $
group element $G_{L}\left( z\right) $ in eq.(\ref{gl}) so that the
discussion given in this section directly applies to the SU$\left(
2/1,1\right) $ model. The most general $g_{L,R}$ can always be parametrized
in the triangular form 
\begin{eqnarray}
g_{L}\left( z\right) &=&e^{it^{+}X^{-}}e^{it_{2}X_{2}}e^{it^{-}X^{+}}=\left( 
\begin{array}{cc}
e^{\frac{1}{2}X_{2}} & -X^{+}e^{\frac{1}{2}X_{2}} \\ 
-X^{-}e^{\frac{1}{2}X_{2}} & e^{-\frac{1}{2}X_{2}}+e^{\frac{1}{2}%
X_{2}}X^{+}X^{-}
\end{array}
\right)  \label{L} \\
g_{R}\left( \bar{z}\right) &=&e^{it^{+}\tilde{X}^{-}}e^{it_{2}\tilde{X}%
_{2}}e^{it^{-}\tilde{X}^{+}}=\left( 
\begin{array}{cc}
e^{\frac{1}{2}\tilde{X}_{2}} & -\tilde{X}^{+}e^{\frac{1}{2}\tilde{X}_{2}} \\ 
-\tilde{X}^{-}e^{\frac{1}{2}\tilde{X}_{2}} & e^{-\frac{1}{2}\tilde{X}%
_{2}}+e^{\frac{1}{2}\tilde{X}_{2}}\tilde{X}^{+}\tilde{X}^{-}
\end{array}
\right)  \label{R}
\end{eqnarray}
By comparing (\ref{figamma}) to (\ref{LR}) and (\ref{L},\ref{R}) one obtains
the general classical solution for the fields $\phi \left( z,\bar{z}\right) $%
, $\gamma ^{-}\left( z,\bar{z}\right) $, $\gamma ^{+}\left( z,\bar{z}\right) 
$ in terms of the left/right moving fields $X^{\pm }\left( z\right)
,X_{2}\left( z\right) $ and $\bar{X}^{\pm }\left( \bar{z}\right) ,\bar{X}%
_{2}\left( \bar{z}\right) $ (see the explicit formulas (\ref{ads11}-\ref
{ads13}) below).

The quantum theory at the conformal critical point also takes the form (\ref
{LR}). The left/right currents are $J\left( z\right) =ik\left( \partial
g\right) g^{-1}=ik\left( \partial g_{L}\right) g_{L}^{-1}$ and $\bar{J}%
\left( \bar{z}\right) =-ikg^{-1}\left( \bar{\partial}g\right)
=-ikg_{R}^{-1}\left( \bar{\partial}g_{R}\right) $. It was shown in \cite
{Bars} \cite{Str95} that the quantum theory parametrized in this form
reduces to a free field theory. The currents and the energy-momentum tensor
are then expressed in terms of canonical sets of free fields: The left
moving sets are $(X^{-}\left( z\right) ,P^{+}\left( z\right) )$ and $\left(
X_{2}\left( z\right) ,P_{2}\left( z\right) \right) $ in terms of which the
left currents $J^{\mu }\left( z\right) $ take the form (after taking into
account normal ordering which shifts the overall $k$ to $k+2$ in the second
equation below) 
\begin{eqnarray}
J^{1}(z)+J^{0}(z) &=&P^{+}(z)  \label{wzwcurrent} \\
J^{1}(z)-J^{0}(z) &=&-:X^{-}\,P^{+}X^{-}:-2P_{2}\,X^{-}+i\left( k+2\right)
\partial _{z}X^{-} \\
J_{2}(z) &=&:X^{-}\,P^{+}:+P_{2}(z)  \label{wzw3}
\end{eqnarray}
where the canonical momenta $P^{+}$ and $P_{2}$ are identified as follows 
\begin{eqnarray}
P^{+}(z) &=&ik\partial _{z}X^{+}e^{X_{2}}  \label{canonical} \\
P_{2}(z) &=&-\frac{1}{2}ik\partial _{z}X_{2}  \nonumber
\end{eqnarray}
Therefore $X^{+}\left( z\right) $ must be expressed in terms of $P^{+}(z)$
as follows 
\begin{equation}
X^{+}\left( z\right) =q^{+}-\frac{i}{k}\int^{z}dz^{\prime
}P^{+}(z^{\prime })e^{-X_{2}\left( z^{\prime }\right) }  \label{X+}
\end{equation}
The Sugawara energy momentum tensor takes the form (after careful ordering
of operators, including zero modes, to insure hermiticity) 
\begin{equation}
T_{++}\left( z\right) =:P^{+}i\partial X^{-}:+\frac{1}{k}\left( :P_{2}^{2}:-%
\frac{i}{z}\partial \left( zP_{2}\right) +\frac{1}{4z^{2}}\right)
\label{T++}
\end{equation}
(and similarly for $\bar{T}_{--}\left( \bar{z}\right) $ ). The $L_{0}$
Virasoro operator that follows from this is 
\begin{equation}
L_{0}=p^{+}p^{-}+\frac{1}{k}\left( \frac{1}{4}+p_{2}^{2}\right) +oscillators
\label{spectr}
\end{equation}
where the zero modes $p^{+},p^{-},p_{2}$ come from 
\begin{eqnarray}
X^{-}\left( z\right) &=&q^{-}-ip^{-}\ln z+oscillators,\quad P^{+}\left(
z\right) =\frac{p^{+}}{z}+oscillators,  \label{xminus} \\
P_{2}\left( z\right) &=&\frac{p_{2}}{z}+oscillators
\end{eqnarray}
This free field representation is similar to Wakimoto's \cite{wakim} (with $%
\gamma \sim X^{-}$ and $\beta \sim P^{+}$) but there are two important
differences:

\begin{itemize}
\item  The fields ($X^{-},P^{+})$ and $\left( X_{2},P_{2}\right) $ are all
hermitian, and similarly all the currents $J^{\mu }$ are hermitian, e.g. the
hermitian conjugate of $J^{1}(z)+J^{0}(z)$ is itself (see details in \cite
{Bars}). This feature is different in SU$\left( 2\right) $ and this is why
we can build a unitary representation with free fields in this
parametrization for SL$\left( 2,R\right) $. By contrast for SU$\left(
2\right) ,$ hermiticity and hence unitarity is not straightforward in the
Wakimoto free field representation and it requires certain screening charges
and singular states etc. to identify the unitary subset of states in the
Hilbert space. None of these occur in SL$\left( 2,R\right) $ and hermiticity
(and unitarity) is manifest at every step in the formulation given above.
Note that $T_{++}\left( z\right) $ and the Virasoro operators $L_{n}$ that
follow from it are also hermitian (i.e. $L_{n}^{\dagger }=-L_{-n}$) thanks
to the hermiticity of the structure $-\frac{i}{z}\partial \left(
zP_{2}\right) +\frac{1}{4z^{2}}$.

\item  The momentum zero mode $p^{-}$ (and similarly $\tilde{p}^{+}$ for
right movers) contributes logarithmic terms to the currents given above (\ref
{wzwcurrent}-\ref{wzw3}). In SU$\left( 2\right) $ these zero modes are
dropped to insure that the currents are periodic for a string when $\sigma
\rightarrow \sigma +2\pi $. However for SL$\left( 2,R\right) $ it is
essential to retain the $p^{-}$ (and $\tilde{p}^{+}$) zero mode because
otherwise the string cannot be put on shell when it is excited as will be
discussed now.
\end{itemize}

If the zero mode $p^{-}$ is dropped then the currents are holomorphic as in
usual affine current algebra (usual affine currents do not have logarithmic
terms $\ln z$ and $\left( \ln z\right) ^{2}$ contributed through the $X^{-}$
in (\ref{xminus})). We define the affine currents as $j^{\mu }\left(
z\right) =\left[ J^{\mu }\left( z\right) \right] _{p^{-}=0}.$ These currents
obey the standard operator products expected from SL$\left( 2,R\right) $
currents. Then the spectrum is given in the form $L_{0}|_{p^{-}=0}=-\frac{1}{%
k}j\left( j+1\right) +oscillators$, a formula that is valid only for affine
currents. Comparing to the expression of $L_{0}$ above we learn that the 
{\it affine current algebra} is realized only in the principal series 
\begin{equation}
j\left( j+1\right) =-\frac{1}{4}-p_{2}^{2},\quad \rightarrow \quad j=-\frac{1%
}{2}+ip_{2}.
\end{equation}
The {\it discrete series or supplementary series do not occur} in this
realization of the affine current algebra since for them $j\left( j+1\right)
>-1/4.$

However there is a physical problem if $p^{-}=0$: since every term is
positive in $L_{0}|_{p^{-}=0}$ it is impossible to put the theory on shell
for any excited level by requiring $L_{0}|_{p^{-}=0}=a\leq 1$ (the
oscillators contribute an integer $\geq 1$ for an excited level, and any
additional {\it spacelike} dimensions contribute only positive terms). This
is why the $p^{-}$ zero mode must be included since $%
p^{+}p^{-}=-p_{0}^{2}+p_{1}^{2}$ is the only negative term in the correct $%
L_{0},$ and its presence enables the theory to be put on-shell, as is also
the case in flat string theory with only one timelike coordinate. The
presence of $p^{-}$ is completely natural as part of $X^{-}$ in the WZW
model, and there would be no issue if one treated the model directly in
terms of the free field representation given above. However a non-zero $%
p^{-} $ modifies certain standard {\it affine} current algebra results and
therefore one must be careful in trying to apply results obtained in affine
current algebra to the SL$\left( 2,R\right) $ WZW string theory as explained
in the next paragraph and in the later sections of this paper. The early
papers until 1995 and some recent papers have overlooked this point. As we
will see below it has consequences for and adds refinements to the AdS-CFT
correspondence.

With a non-zero $p^{-}$ the current algebra {\it is not the usual affine
current algebra} since it includes the logarithmic terms $\ln z$ and $\left(
\ln z\right) ^{2}$. In terms of the affine currents $j^{\mu }\left( z\right) 
$ the correct SL$\left( 2,R\right) $ WZW currents $J^{\mu }\left( z\right) $
are given by \cite{Bars} \cite{Str95} 
\[
\begin{array}{l}
J^{0}(z)+J^{1}(z)=\left[ j^{0}(z)+j^{1}(z)\right] , \\ 
J^{0}(z)-J^{1}(z)=\left[ j^{0}(z)-j^{1}(z)\right] -2ip^{-}\ln z\,\,j^{2}(z)
\\ 
\quad \quad -\frac{k+2}{z}p^{-}+\left( -ip^{-}\ln z\right) ^{2}\left[
j^{0}(z)+j^{1}(z)\right] , \\ 
J^{2}(z)=j^{2}(z)-ip^{-}\ln z\,\left[ j^{0}(z)+j^{1}(z)\right] .
\end{array}
\]
Despite the logarithms, The Sugawara form $T\sim JJ$ gives the correct
energy momentum tensor in (\ref{T++}), and the WZW currents $J^{\mu }\left(
z\right) $ have the correct operator products among themselves and with the
energy-momentum tensor as shown in detail in \cite{Bars} \cite{Str95}.

The remaining problem is the periodicity of the currents $J^{\mu }\left(
ze^{in2\pi }\right) =J^{\mu }\left( z\right) $, which is not true in
general, but must be true at least in the physical sector of the theory. It
was shown that this condition imposes a quantization condition on the zero
modes \cite{Bars} \cite{Str95} 
\begin{equation}
p^{+}p^{-}=-r,\quad r\in Z_{+}
\end{equation}
where $r$ is a {\it positive} integer or zero. Therefore the physical states
must come in sectors labeled by $r$. The spectrum of the SL$\left(
2,R\right) $ WZW model may now be written as 
\begin{equation}
L_{0}=-r-\frac{j\left( j+1\right) }{k}+{\it {integer}=a\leq 1}
\end{equation}
where $j=-1/2+is$ ($s$ is the eigenvalue of $p_{2})$ labels the
representation of the affine current algebra $j^{\mu }\left( z\right) $ and $%
-r$ comes purely through the extra zero mode $p^{-}$. As argued below, $r$
is related to winding modes at the AdS boundary and therefore represents
winding strings.

The right moving currents $J_R=-ikg_R^{-1}\left( \partial g_R\right) $ are
quantized in a similar way. In this case the independent canonical degrees
of freedom are $\left( \tilde{X}^{+}\left( \bar{z}\right) ,\tilde{P}%
^{-}\left( \bar{z}\right) \right) $ and $\left( \tilde{X}_2\left( \bar{z}%
\right) ,\tilde{P}_2\left( \bar{z}\right) \right) $, and $\tilde{X}%
^{-}\left( \bar{z}\right) $ is expressed in terms of $\tilde{P}^{-}$%
\begin{equation}
\tilde{X}^{-}\left( \bar{z}\right) =\tilde{q}^{-}-
\frac ik\int^{\bar{
z}}d\bar{z}^{\prime }\tilde{P}^{-}(\bar{z}^{\prime })\,e^{-\tilde{X}_2\left( 
\bar{z}^{\prime }\right) }  \label{X-}
\end{equation}
while the currents are 
\begin{eqnarray}
\tilde{J}^1(\bar{z})-\tilde{J}^0(\bar{z}) &=&-\tilde{P}^{-}(\bar{z}) \\
\tilde{J}^1(\bar{z})+\tilde{J}^0(\bar{z}) &=&:\tilde{X}^{+}\,\tilde{P}^{-}\,%
\tilde{X}^{+}:+2\tilde{P}_2\,\tilde{X}^{+}-i\left( k+2\right) \partial _{%
\bar{z}}\tilde{X}^{+}(\bar{z}) \\
\tilde{J}_2(\bar{z}) &=&-:\tilde{X}^{+}\,\tilde{P}^{-}:-\tilde{P}_2(\bar{z})
\end{eqnarray}
The spectrum of right movers is given by 
\begin{equation}
\tilde{L}_0=\tilde{p}^{+}\tilde{p}^{-}+\frac 1k\left( \frac 14+\tilde{p}%
_2^2\right) +oscillators
\end{equation}

Combining the left and right movers and using the conditions $j=\tilde{j}$
and $L_{0}=\tilde{L}_{0}$ we find $p_{2}=\tilde{p}_{2}$ and 
\begin{equation}
\tilde{p}^{+}\tilde{p}^{-}=p^{+}p^{-}=-r,\quad r\in Z_{+}\,.
\label{monodrom}
\end{equation}
Therefore physical states are constructed by applying oscillators to the
base that is labeled by 
\begin{equation}
|base>=|p^{+},p^{-},p_{2};\tilde{p}^{+},\tilde{p}^{-},\tilde{p}_{2}>
\label{base}
\end{equation}
with the relations and quantization conditions given above. The Virasoro
constraints are applied to single out the physical states. The no ghost
theorem was proven in \cite{Bars} \cite{Str95}.

The procedure above establishes the physical spectrum of the string on AdS$%
_{3}$. The next step is to construct vertex operators that correspond to
these states.

\section{Vertex operator}

Vertex operators are constructed by starting with the ``tachyon'' vertex
operator, which is the group element in a representation as specified by the
generators $\hat{t}^{\mu }$ (the relevant representation will be specified
below) 
\begin{eqnarray}
V\left( g\right) &=&e^{i\gamma ^{-}\hat{t}^{+}}e^{i\phi \hat{t}%
_{2}}e^{i\gamma ^{+}\hat{t}^{-}}  \label{vertexfull} \\
&=&V\left( g_{L}\right) V\left( g_{R}\right)
\end{eqnarray}
The second line $V\left( g\right) =V\left( g_{L}\right) V\left( g_{R}\right) 
$ follows from the group property in any representation. Furthermore, $%
V\left( g_{L}\right) $ and $V\left( g_{R}\right) $ are constructed from free
fields $X^{\mu }\left( z\right) $ and $\bar{X}^{\mu }\left( \bar{z}\right) $
given above 
\begin{equation}
V\left( g_{L}\right) =e^{iX^{-}\hat{t}^{+}}e^{iX_{2}\hat{t}_{2}}e^{iX^{+}%
\hat{t}^{-}},\quad V\left( g_{R}\right) =e^{i\tilde{X}^{-}\hat{t}^{+}}e^{i%
\tilde{X}_{2}\hat{t}_{2}}e^{i\tilde{X}^{+}\hat{t}^{-}}  \label{vertexLR}
\end{equation}
The representation of the generators $\hat{t}^{\mu }$ should correspond to
the physical states determined above (as opposed to the non-unitary $j=1/2$
representation of footnote 1). We need to take into account the effect of
the zero mode $p^{-}$ in the representation $\hat{t}^{\mu }$ so that there
is operator-state correspondence. The relevant representation will be
discussed in the next subsection.

In the quantum theory the expressions for$\ V\left( g_{L,R}\right) $ must be
normal ordered appropriately so that their operator products with the
currents and stress tensor give the correct results for the single and
double poles. An additional desirable property is that the AdS vertex
operators should tend to the flat 3D string vertex operators $exp\left(
ik_{\mu }X^{\mu }\left( z,\bar{z}\right) \right) $ since the large $k$ limit
is smooth for the complete theory as we have already discussed. We have
accomplished all of these properties as described below. Using this
construction in principle one can perform computations of correlation
functions using free fields.

For semi-classical arguments used in the interpretation of the model, it is
also useful to express the AdS coordinates $\phi \left( z,\bar{z}\right) $, $%
\gamma ^{+}\left( z,\bar{z}\right) $, $\gamma ^{-}\left( z,\bar{z}\right) $
themselves in terms of the free fields. If we ignore orders of operators in
a semi-classical approach, we can obtain the result by computing $V\left(
g_{L}\right) V\left( g_{R}\right) $ in the $j=1/2$ representation. Thus,
using (\ref{LR}, \ref{figamma}) and (\ref{L}, \ref{R}) we obtain

\begin{eqnarray}
\phi \left( z,\bar{z}\right) &=&X_2\left( z\right) +\tilde{X}_2\left( \bar{z}%
\right) +2\ln \left( 1+X^{+}\left( z\right) \tilde{X}^{-}\left( \bar{z}%
\right) \right)  \label{ads11} \\
\gamma ^{-}\left( z,\bar{z}\right) &=&X^{-}\left( z\right) +\frac{%
e^{-X_2\left( z\right) }\tilde{X}^{-}\left( \bar{z}\right) }{1+X^{+}\left(
z\right) \tilde{X}^{-}\left( \bar{z}\right) }  \label{ads12} \\
\gamma ^{+}\left( z,\bar{z}\right) &=&\tilde{X}^{+}\left( \bar{z}\right) +%
\frac{e^{-\tilde{X}_2\left( \bar{z}\right) }X^{+}\left( z\right) }{%
1+X^{+}\left( z\right) \tilde{X}^{-}\left( \bar{z}\right) }  \label{ads13}
\end{eqnarray}
where $X^{+}\left( z\right) ,\tilde{X}^{-}\left( \bar{z}\right) $ are given
by (\ref{X+}) and (\ref{X-}) 
\begin{eqnarray}
X^{+}\left( z\right) &=&q^{+}-\frac ik\int^zdz^{\prime
}P^{+}(z^{\prime })e^{-X_2\left( z^{\prime }\right) },\quad \\
\tilde{X}^{-}\left( \bar{z}\right) &=&\tilde{q}^{-}-\frac ik\int^{%
\bar{z}}d\bar{z}^{\prime }\tilde{P}^{-}(\bar{z}^{\prime })\,e^{-\tilde{X}%
_2\left( \bar{z}^{\prime }\right) }.
\end{eqnarray}
We will use these expressions below to discuss windings at the AdS boundary $%
\phi \rightarrow \infty $.

\subsection{Position-momentum basis and state-operator correspondence}

The following is an operator representation of the SL$\left( 2,R\right) $
generators $\hat{t}^{\mu }$ that correspond to the physical states
determined in the previous section\footnote{%
This unitary representation is obtained by considering the quantum theory of
a particle in AdS$_{3}$ space. In momentum space its complete normalizable
wavefunctions can be related to those of the Hydrogen atom in an appropriate
basis. For an informal discussion see \cite{sl2rpart}.}. They should be
inserted into the expression of the vertex operators given in (\ref
{vertexfull}-\ref{vertexLR})

\begin{equation}
\begin{array}{lll}
\hat{t}^{+} & \equiv \hat{p}^{+} & \equiv -\hat{x}^{+}\hat{p}^{-}\hat{x}%
^{+}+2s\hat{x}^{+}-\frac{kr}{\hat{p}^{-}} \\ 
\hat{t}_{2} & \equiv \frac{1}{2}\left( \hat{x}^{-}\hat{p}^{+}+\hat{p}^{+}%
\hat{x}^{-}\right) +s & \equiv -\frac{1}{2}\left( \hat{x}^{+}\hat{p}^{-}+%
\hat{p}^{-}\hat{x}^{+}\right) +s \\ 
\hat{t}^{-} & \equiv -\hat{x}^{-}\hat{p}^{+}\hat{x}^{-}-2s\hat{x}^{-}-\frac{%
kr}{\hat{p}^{+}} & \equiv \hat{p}^{-}
\end{array}
\label{trep}
\end{equation}
These forms may look familiar except for the terms that contain $kr$. The
integer $r$ corresponds to the quantum label of the state (\ref{base}) as
determined by the monodromy argument (\ref{monodrom}), and $k$ is the
central extension of the current algebra. The operators $\left( \hat{x}^{-},%
\hat{p}^{+}\right) $ form a canonical pair. These $\hat{x}^{-},\hat{p}^{+}$
operators are just a convenient device, they are {\it not} the zero modes of
the fields $X^{-},P^{+}$ etc. Similarly the pair $\left( \hat{x}^{+},\hat{p}%
^{-}\right) $ is canonical, but it is not independent of the pair $\left( 
\hat{x}^{-},\hat{p}^{+}\right) $. The relationship between these pairs is
non-linear and is given by the two forms of the generators above. The
structure of $\hat{t}^{\mu }$ given above is obtained by studying the
particle moving in an AdS space and it is interesting that it can represent
all possible representations of SL$\left( 2,R\right) $ by taking various
values of $s$ and $kr.$

By using the canonical commutation rules $\left[ \hat{x}^{-},\hat{p}^{+}%
\right] =\left[ \hat{x}^{-},\hat{p}^{+}\right] =i$ it is easy to see that
either form of the operators $\hat{t}^\mu $ satisfy the same commutation
rules as the 2$\times 2$ matrix representation of footnote 1. Furthermore
the $\hat{t}^\mu $ are manifestly hermitian thus insuring that they
correspond to a unitary representation (unlike the $t^\mu $ of footnote 1)$.$
The Casimir operator for either form of $\hat{t}^\mu $ is 
\begin{eqnarray}
\hat{C}_2 &=&-\frac 12\left( \hat{t}^{+}\hat{t}^{-}+\hat{t}^{-}\hat{t}%
^{+}\right) -\hat{t}_2^2  \label{casim} \\
&=&kr-\frac 14-s^2
\end{eqnarray}
(the corresponding Casimir eigenvalue for the $t^\mu $ of footnote 1 is $%
C_2=3/4$). Observe that $-\frac 1k\hat{C}_2=-r+\frac 1k\left( \frac
14+s^2\right) $ is the same as the formula for the spectrum of a physical
state as given in (\ref{spectr}). We will see that the operator product of
the energy-momentum tensor with the vertex operator gives 
\begin{eqnarray}
T\left( z\right) V^{r,s}\left( w\right) &\sim &\frac{-r+\frac 1k\left( \frac
14+s^2\right) }{\left( z-w\right) ^2}V^{r,s}\left( z\right) +\frac{\partial
V^{r,s}\left( z\right) }{\left( z-w\right) }  \label{weightsrs} \\
\Delta ^{r,s} &=&-r+\frac 1k\left( \frac 14+s^2\right)
\end{eqnarray}
where $\Delta ^{r,s}$ is the conformal dimension. This identifies $\hat{t}%
^\mu $ given above as the representation that provides the desired {\it %
state-operator correspondence}. Note that the contribution of the zero modes
are taken into account through the term $kr$. Without the $kr$ term one
cannot construct primary vertex operators of the form $:J^\mu V^{r,s}:$
(needed for the AdS-CFT correspondence) with total conformal weight $1,$
since this requires $\Delta ^{r,s}=0$.

There is here the possibility for some confusion about this representation
and we would like to comment on it. Of course we may define a $\hat{\jmath}$
through 
\begin{equation}
\hat{C}_{2}=kr-\frac{1}{4}-s^{2}=\hat{\jmath}\left( \hat{\jmath}+1\right)
\end{equation}
and note that we must have $\hat{\jmath}\left( \hat{\jmath}+1\right) >0$ to
be able to satisfy the mass shell condition. This identifies the $\hat{t}%
^{\mu }$ in the discrete series representation. How is this possible since
we made the point that only the principal series representation for the
affine currents is allowed? The answer is that the $\hat{t}^{\mu }$ takes
into account the contribution of the zero mode while the affine currents
continue to be in the principal series. That is, the full {\it current
algebra module}, including the string excitations, is in the principal
series of the affine currents $j^{\mu }\left( z\right) $, which is a very
different module than the discrete series module. When this is combined with
the zero mode, the excitation spectrum of the current algebra module remains
the same, but the mass shell condition changes, and this is taken into
account in the WZW vertex operators constructed with $\hat{t}^{\mu }.$ The
affine current algebra module is still in the principal series, but the
physics is supplemented by the zero modes. The zero mode is related to
winding strings at the AdS boundary. This shows that the effects of the zero
modes are subtle, as we will continue to witness further in the following
sections.

The group theoretical states for SL$\left( 2,R\right) $ are usually labeled
as $|jm>$ where $m$ is the eigenvalue of the compact generator $\hat{t}^{0}.$
One may consider states in which other operators are diagonal. For our
purposes it is useful to diagonalize $\hat{t}^{+}$ or $\hat{t}^{-}$. When $%
\hat{t}^{+}$ is diagonal in the basis $<s,p^{+}|$, it corresponds to
diagonalizing the operator $\hat{p}^{+}$ and when $\hat{t}^{-}$ is diagonal
in the basis $|s,p^{-}>$, it corresponds to diagonalizing the operator $\hat{%
p}^{-}$. The Fourier transform of these states correspond to diagonalizing
the operators $\hat{x}^{-}$ or $\hat{x}^{+}$ in the basis $<s,x^{-}|$ or $%
|s,x^{+}>$, respectively. In position space we have $<s,x^{-}|\hat{p}^{+}=-i%
\frac{\partial }{\partial x^{-}}<s,x^{-}|$ and $\hat{p}^{-}|s,x^{+}>=i\frac{%
\partial }{\partial x^{+}}|s,x^{+}>$ consistent with the commutation rules.

One may evaluate the matrix elements of the vertex operators (\ref
{vertexfull}-\ref{vertexLR}) in the position or momentum basis in the same
way that one would compute the group elements in the $|jm>$ basis $%
D_{mm^{\prime }}^j\left( g\right) $. In particular we find it convenient to
use the momentum basis or the position basis as follows 
\begin{eqnarray}
V_{p^{+},p^{-}}^{r,s}\left( g\right) &=&<p^{+}|e^{i\gamma ^{-}\hat{t}%
^{+}}e^{i\phi \hat{t}_2}e^{i\gamma ^{+}\hat{t}^{-}}|p^{-}> \\
V_{x^{-},x^{+}}^{r,s}\left( g\right) &=&<x^{-}|e^{i\gamma ^{-}\hat{t}%
^{+}}e^{i\phi \hat{t}_2}e^{i\gamma ^{+}\hat{t}^{-}}|x^{+}>
\end{eqnarray}
and similarly for the left/right vertex operators. Note that in momentum
basis $\hat{t}^{+}$ is diagonal on the bra and the $\hat{t}^{-}$ is diagonal
on the ket, showing that this form is already fairly close to the vertex
operator of a flat string. We will see that these labels $p^{\pm }$ or $%
x^{\pm }$ have the interpretation of momenta or positions at the boundary of
AdS$_3.$ The position basis will be useful to connect to the discussions in
the recent literature and show the refinements that need to be made. The
momentum basis will be useful for establishing the operator product (\ref
{weightsrs}) at the fully quantum level (ordering of operators taken into
account) and for discussing the flat limit as $k\rightarrow \infty .$ Our
new vertex operator has the desirable properties.

\subsection{Vertex operator in position basis}

The vertex operator in position space is defined by

\begin{eqnarray}
V_{x^{-},x^{+}}^{r,s}\left( g\right) &=&<x^{-}|e^{i\gamma ^{-}\hat{t}%
^{+}}e^{i\phi \hat{t}_{2}}e^{i\gamma ^{+}\hat{t}^{-}}|x^{+}> \\
&=&<\left( x^{-}+\gamma ^{-}\right) |e^{i\phi \hat{t}_{2}}|\left(
x^{+}-\gamma ^{+}\right) > \\
&=&<e^{\phi }\left( x^{-}+\gamma ^{-}\right) |\left( x^{+}-\gamma
^{+}\right) >e^{i\left( s-i/2\right) \phi } \\
&=&<\left( x^{-}+\gamma ^{-}\right) |e^{\phi }\left( x^{+}-\gamma
^{+}\right) >e^{i\left( s-i/2\right) \phi }
\end{eqnarray}
where we have used that $\hat{t}^{+}=\hat{p}^{+}$ and $\hat{t}^{-}=\hat{p}%
^{-}$ are translation operators on $x^{\pm }$ space, and $\hat{t}_{2}$ is a
dilation operator on $x^{\pm }$ space (except for the extra factor $%
e^{i\left( s-i/2\right) \phi }$). If we define $f_{r,s}\left(
x^{-}x^{+}\right) \equiv <x^{-}|x^{+}>$ then we must have a function of only
the product $x^{-}x^{+}$ on account of its properties under dilations. Then
the full vertex is 
\begin{equation}
V_{x^{-},x^{+}}^{r,s}\left( g\right) =e^{i\left( s-i/2\right) \phi
}f_{r,s}\left( e^{\phi }\left( x^{+}-\gamma ^{+}\right) \left( x^{-}+\gamma
^{-}\right) \right) .
\end{equation}
To find $f_{r,s}\left( x^{-}x^{+}\right) $ sandwich $\hat{t}^{-}$ and use
its action on the right and left states to obtain the differential equation 
\begin{eqnarray}
&<&x^{-}|\hat{t}^{-}|x^{+}>=i\partial _{+}f_{r,s} \\
&=&ix^{-}\partial _{-}\left( x^{-}f_{r,s}\right) -2sx^{-}f_{r,s}-\frac{kr}{%
-i\partial _{-}}f_{r,s}
\end{eqnarray}
At $r=0$ this first order differential equation has the unique solution $%
f_{0,s}=\sqrt{\frac{1}{\pi }}\left( x^{+}x^{-}+1\right) ^{-1-2is}$, and the
full vertex operator at $r=0$ becomes 
\begin{equation}
V_{x^{-},x^{+}}^{0,s}\left( g\right) =\sqrt{\frac{1}{\pi }}\left( e^{\phi
/2}\left( x^{+}-\gamma ^{+}\right) \left( x^{-}+\gamma ^{-}\right) +e^{-\phi
/2}\right) ^{-1-2is}.  \label{V0s}
\end{equation}
At non-zero values of $r$ we apply $-i\partial _{-}$ on both sides to obtain
the second order differential equation 
\begin{equation}
-\partial _{-}\left( x^{-}\partial _{-}\left( x^{-}f_{r,s}\right) \right)
-2is\partial _{-}\left( x^{-}f_{r,s}\right) +krf_{r,s}+\partial _{-}\partial
_{+}f_{r,s}=0
\end{equation}
This becomes the hypergeometric equation in one variable. The exact solution
that is well behaved at infinity is 
\begin{eqnarray}
f_{r,s}\left( x^{-}x^{+}\right) &=&\frac{\left( 1+x^{-}x^{+}\right) ^{-1-2is}%
}{\sqrt{\pi }\left( x^{-}x^{+}\right) ^{\sigma -is}}\,_{2}F_{1}\left( \sigma
-is,\sigma -is;1+2\sigma ;\frac{-1}{x^{-}x^{+}}\right) , \\
\sigma &\equiv &\sqrt{kr-s^{2}}
\end{eqnarray}
When $r=0$, we get $\sigma =is$ and $_{2}F_{1}\left( 0,0;1+2is;\frac{-1}{%
x^{-}x^{+}}\right) =1$, and the solution reduces to $V_{x^{-},x^{+}}^{0,s}%
\left( g\right) $ given above. For general $r$, with $kr>s^{2}$ that
satisfies the mass shell condition for excited strings, the vertex operator
becomes 
\begin{eqnarray}
V_{x^{-},x^{+}}^{r,s}\left( g\right) &=&\sqrt{\frac{1}{\pi }}\left( e^{\phi
/2}\left( x^{+}-\gamma ^{+}\right) \left( x^{-}+\gamma ^{-}\right) +e^{-\phi
/2}\right) ^{-1-2is}\times  \label{vertexrs} \\
&&\times \frac{_{2}F_{1}\left( \sigma -is,\sigma -is;1+2\sigma ;\frac{-1}{%
\left( x^{+}-\gamma ^{+}\right) \left( x^{-}+\gamma ^{-}\right) e^{\phi }}%
\right) }{\left( \left( x^{+}-\gamma ^{+}\right) \left( x^{-}+\gamma
^{-}\right) e^{\phi }\right) ^{\sigma -is}}  \nonumber
\end{eqnarray}
We see that the zero mode modifies the vertex operator by the factor in the
second line. The conformal weight of this vertex operator is 
\begin{equation}
\Delta ^{r,s}=-r+\frac{1}{k}\left( \frac{1}{4}+s^{2}\right)
\end{equation}
as in (\ref{weightsrs}) which will be proven below at the full quantum level.

The factor in the second line of (\ref{vertexrs}) has been missed in recent
discussions of the AdS-CFT correspondence. The vertex operator used in
recent literature either is not unitary \cite{GKS}-\cite{kutseib}, or as in 
\cite{teschner} corresponds to only the first line of (\ref{vertexrs}),
which is the vertex operator at $r=0$. As already emphasized several times,
with $r=0$ the mass shell condition, or the primary operator condition $%
\left( {\rm {with }\Delta ^{r,s}\leq 0}\right) $ could not be satisfied and
therefore the state-operator correspondence and/or unitarity would be lost.
Our new vertex operator has the desirable properties of unitarity and
state-operator correspondence.

\subsubsection{Vertex operator near the AdS$_{3}$ boundary}

The behavior of the correct vertex operator near the AdS boundary $\phi
\rightarrow \infty $ can now be examined. As it turns out the important
quantity is not the vertex operator, which is the wavefunction, but rather
the probability, that involves the absolute value square of the wavefunction 
\begin{equation}
\Omega _{x^{-},x^{+}}^{r,s}\left( g\right) =\left|
V_{x^{-},x^{+}}^{r,s}\left( g\right) \right| ^{2}.
\end{equation}
Noting that $_{2}F_{1}\left( a,b;c;0\right) =1$, we see that as long as $%
\left( x^{+}-\gamma ^{+}\right) \left( x^{-}+\gamma ^{-}\right) \neq 0$, the
vertex operator falls off as $\exp \left( -\phi \left( \frac{1}{2}+\sqrt{%
kr-s^{2}}\right) \right) ,$ or $\Omega _{x^{-},x^{+}}^{r,s}\left( g\right)
\rightarrow \exp \left( -\phi \left( 1+2\sqrt{kr-s^{2}}\right) \right) $,
indicating that this wavefunction is normalizable. When $\left( x^{+}-\gamma
^{+}\right) \left( x^{-}+\gamma ^{-}\right) \sim 0$ the first factor in (\ref
{vertexrs}) behaves like a delta function as $\phi \rightarrow \infty $,
while the second factor modifies it mildly with logarithms 
\begin{equation}
\Omega _{x^{-},x^{+}}^{r,s}\rightarrow \delta \left( x^{+}+\gamma
^{+}\right) \delta \left( x^{-}-\gamma ^{-}\right) \times \left[ c_{1}\ln
\left( x^{+}-\gamma ^{+}\right) \left( x^{-}+\gamma ^{-}\right) +c_{1}\phi
+c_{2}\right] ^{2}
\end{equation}
where $c_{1,2}$ are independent of $(x^{+}-\gamma ^{+})$, $(x^{-}+\gamma
^{-})$. In ariving at this result we have used the small $z\rightarrow 0$
behavior of the hypergeometric function 
\begin{equation}
\frac{_{2}F_{1}\left( \sigma -is,\sigma -is;1+2\sigma ;\frac{-1}{z}\right) }{%
z^{\sigma -is}}\rightarrow \frac{\Gamma \left( 1+2\sigma \right) \times \ln z%
}{\Gamma \left( \sigma -is\right) \Gamma \left( 1+\sigma +is\right) }%
+c_{2}\left( \sigma ,s\right)
\end{equation}
Note that at $r=0$ we get $\sigma =is$ and the coefficients $c_{1}=0$ and $%
c_{2}=1.$ When $r\neq 0$ the term $c_{1}\phi +c_{2}$ is neglected as
compared to the first term in $c_{1}\ln \left( x^{+}-\gamma ^{+}\right)
\left( x^{-}+\gamma ^{-}\right) +\left( c_{1}\phi +c_{2}\right) $ since $%
\left( x^{+}-\gamma ^{+}\right) \left( x^{-}+\gamma ^{-}\right) \sim 0$
before taking the limit $\phi \rightarrow \infty .$ So we see that the
support of the probability at the boundary of AdS $\phi \rightarrow \infty $
is precisely at 
\begin{equation}
\gamma ^{+}\left( z,\bar{z}\right) =x^{+},\,\quad \gamma ^{-}\left( z,\bar{z}%
\right) =-x^{-}.
\end{equation}
Therefore, the labels $x^{\pm }$ must be interpreted as the coordinates at
the boundary of the AdS space, in agreement with \cite{ooguri}, \cite
{teschner}, for any $r.$

\subsubsection{Windings at the AdS boundary}

To examine further the properties of the string theory near the boundary at $%
\phi \rightarrow \infty $, in particular its zero modes, we take the center
of mass positions of $X_{2},\tilde{X}_{2}$ to infinity. That is, use $%
X_{2}\left( z\right) =q_{2}+\cdots $ and $\tilde{X}_{2}\left( \bar{z}\right)
=\tilde{q}_{2}+\cdots $ , and let $q_{2}=Q+q$ and $\tilde{q}_{2}=Q-q$ and
then let $Q\rightarrow \infty $. Then note that $\gamma ^{\pm }\left( z,\bar{%
z}\right) $ given in (\ref{ads12},\ref{ads13}) become purely left or right
moving in this limit 
\begin{equation}
\gamma ^{-}\left( z,\bar{z}\right) \rightarrow X^{-}\left( z\right) ,\quad
\gamma ^{+}\left( z,\bar{z}\right) \rightarrow \tilde{X}^{+}\left( \bar{z}%
\right) .
\end{equation}
Then using the Minkowski signature on the worldsheet replace $z\rightarrow
\exp \left( i\left( \tau +\sigma \right) \right) $ and $\bar{z}\rightarrow
\exp \left( i\left( \tau -\sigma \right) \right) $ 
\begin{eqnarray}
X^{-}\left( \tau +\sigma \right) &=&q^{-}+p^{-}\left( \tau +\sigma \right)
+oscillators,\quad P^{+}\left( z\right) =p^{+}+oscl., \\
\tilde{X}^{+}\left( \tau -\sigma \right) &=&\tilde{q}^{+}+\tilde{p}%
^{+}\left( \tau -\sigma \right) +oscillators,\quad \tilde{P}^{-}\left(
z\right) =\tilde{p}^{-}+oscl.,
\end{eqnarray}
and examine the periodicity of the AdS string as $\sigma \rightarrow \sigma
+2\pi $. The oscillator part is periodic; for full periodicity we must make
a periodic lattice in $q^{-}$ space with lattice size $R$ and identify
points on this lattice 
\begin{equation}
q^{-}\sim q^{-}+2\pi p^{-},\quad p^{-}=nR
\end{equation}
Then the canonical conjugate $p^{+}=m/R$ must be quantized in units of $1/R$%
. Thus, we learn that the string winds $n$ times and that $p^{-}$ is
quantized in terms of winding number $n$ while $p^{+}$ is quantized in terms
of the Kaluza-Klein quantum number $m$, and similarly for $\tilde{p}^{-},%
\tilde{p}^{+}$ 
\begin{equation}
p^{-}=nR,\quad p^{+}=\frac{m}{R},\quad \tilde{p}^{+}=\tilde{n}R,\quad \tilde{%
p}^{-}=\frac{\tilde{m}}{R},
\end{equation}
This observation is related to the winding of the long strings discussed in 
\cite{GKS}, \cite{kutseib}, \cite{sw}, so we can identify the winding
numbers 
\begin{equation}
n=\oint \frac{d\gamma ^{-}}{\gamma ^{-}},\quad \tilde{n}=\oint \frac{d\gamma
^{+}}{\gamma ^{+}}.
\end{equation}
Now we can compute the products $p^{+}p^{-},\,\tilde{p}^{+}\tilde{p}^{-}$
that appear in the spectrum (\ref{monodrom}) and establish the following
relation between these integers 
\begin{equation}
mn=\tilde{m}\tilde{n}=-r,\quad r\in Z_{+}
\end{equation}
Thus, the winding of the long strings on the boundary as discussed in \cite
{GKS}, \cite{kutseib}, \cite{sw} demands that $p^{-}=nR$ be non-zero. Of
course, this is in agreement with the requirements of on shell and monodromy 
\cite{Bars}\cite{Str95} we emphasized above. We now see that the effect of
the zero mode $p^{-}$ must be included in the vertex operator if it is to
describe a string that has non-trivial windings in AdS space.

\subsection{Vertex operator in momentum basis}

We will now take advantage of the factorized form of the vertex operator.
The full vertex operator is 
\begin{eqnarray}
V_{p^{+},p^{-}}^{r,s}\left( z,\bar{z}\right) &=&<p^{+}|V\left( g_L\right)
V\left( g_R\right) |p^{-}> \\
&=&\int d\tilde{p}^{-}<p^{+}|V\left( g_L\right) |\tilde{p}^{-}><\tilde{p}%
^{-}|V\left( g_R\right) |p^{-}> \\
&=&\int d\tilde{p}^{-}\,V_{p^{+},\tilde{p}^{-}}^{r,s}\left( z\right) \tilde{V%
}_{\tilde{p}^{-},p^{-}}^{r,s}\left( \bar{z}\right)
\end{eqnarray}
In this section we are going to describe only the left moving part of the
factorized vertex operator in the momentum basis $V_{p^{+},p^{-}}^{r,s}%
\left( z\right) $ and verify that it has the correct operator product
properties with the left moving currents and the stress tensor. The right
moving factor of the factorized vertex operator $\tilde{V}_{\tilde{p}%
^{-},p^{-}}^{r,s}\left( \bar{z}\right) $ is insensitive to these operator
products and therefore we do not include it in the discussion. To discuss
operator products for right movers we insert the intermediate states $1=\int
d\tilde{p}^{+}\,|\tilde{p}^{+}><\tilde{p}^{+}|$ and then the discussion for
right movers parallels the discussion for left movers. After defining the
left moving vertex operator $V_{p^{+},p^{-}}^{r,s}\left( z\right) $ at the
fully quantum level (ordering of operators taken into account) we are going
to show that it has correct quantum operator products with the currents and
that it has the desired conformal dimension 
\[
\Delta ^{r,s}=-r+\frac 1k\left( \frac 14+s^2\right) . 
\]

\subsubsection{Classical expression}

The left moving part of the vertex operator in momentum space is defined by 
\begin{eqnarray}
V_{p^{+},p^{-}}^{r,s}\left( g_L\left( z\right) \right) &=&<s,p^{+}|e^{iX^{-}%
\hat{t}^{+}}e^{iX_2\hat{t}_2}e^{iX^{+}\hat{t}^{-}}|s,p^{-}> \\
&=&e^{iX^{-}p^{+}}<s,p^{+}|e^{iX_2\hat{t}_2}|s,p^{-}>e^{iX^{+}p^{-}} \\
&=&e^{iX^{-}p^{+}}\,\,e^{-\frac 12X_2\left( 1-2is\right)
}<s,e^{-X_2}p^{+}|s,p^{-}>e^{iX^{+}p^{-}} \\
&=&e^{iX^{-}p^{+}}\,\,e^{-\frac 12X_2\left( 1-2is\right)
}<s,p^{+}|s,e^{-X_2}p^{-}>e^{iX^{+}p^{-}}
\end{eqnarray}
where we have used the fact that $\hat{t}^{+}$ is diagonal on $<s,p^{+}|$
and $\hat{t}^{-}$ is diagonal on $|s,p^{-}>$ and that the operator $\hat{t}%
^2 $ is the dilation operator on functions of $p^{\pm }$ except for the
additional overall phase. If we define the function $F_{r,s}\left(
p^{+}p^{-}\right) =<s,p^{+}|s,p^{-}>$ then we must have a function of the
single variable $p^{+}p^{-}$ on account of its properties under dilations.
Then the vertex operator is 
\begin{equation}
V_{p^{+},p^{-}}^{r,s}\left( z\right) =e^{iX^{-}p^{+}}\,\,e^{-\frac
12X_2\left( 1-2is\right) }\,F_{r,s}\left( e^{-X_2}p^{-}p^{+}\right)
e^{iX^{+}p^{-}}  \label{VFrs}
\end{equation}
To find the function $F_{r,s}$ we sandwich $<s,p^{+}|\hat{t}^{\mp }|s,p^{-}>$
and derive a differential equation by operating on both sides 
\begin{equation}
\left( p^{\mp }-\partial _{\pm }p^{\pm }\partial _{\pm }+2is\partial _{\pm }+%
\frac{kr}{p^{\pm }}\right) F_{r,s}(p^{+},p^{-})=0.
\end{equation}
By multiplying through with $p^{\pm }$ this becomes a single equation in one
variable $\kappa =p^{+}p^{-}$ 
\begin{equation}
\left( \kappa -\left( \kappa \partial _\kappa \right) ^2+2is\left( \kappa
\partial _\kappa \right) +kr\right) F_{r,s}(\kappa )=0  \label{Frs}
\end{equation}
The solution is given in terms of a Bessel function 
\begin{equation}
F_{r,s}(\kappa )=\kappa ^{is}\,\,J_{2\sigma }\left( 2\sqrt{-\kappa }\right)
,\quad \sigma =\sqrt{kr-s^2}.
\end{equation}
Therefore the middle factor in the vertex operator is obtained by rescaling $%
\kappa $ with the factor $e^{-X_2}$. The result is 
\begin{equation}
V_{p^{+},p^{-}}^{r,s}\left( z\right) =Ce^{iX^{-}p^{+}}\,\,e^{-\frac
12X_2}\,J_{2\sigma }\left( 2\sqrt{-p^{+}p^{-}}e^{-\frac 12X_2(z)}\right)
\,e^{iX^{+}p^{-}}  \label{braket}
\end{equation}
where $s,p^{\pm }$ have been absorbed into an overall factor $C\left(
r,s,p^{\pm }\right) =C(s,r)\,\times (p^{+}p^{-})^{is}$.

\subsubsection{Quantum ordering and operator products}

At the full quantum level (with an ordering of the operators that will be
given below) the correct operator products with the currents are given as 
\begin{equation}
J^\mu (z)\times V_{p^{+}p^{-}}^{r,s}(w)=\frac 1{z-w}<s,p^{+}|\hat{t}^\mu
V^{r,s}(w)|s,p^{-}>  \label{curvertex}
\end{equation}
where $\mu $ is $+,-$ or 2. The action of $\hat{t}^\mu $ on the bra is a
differential operator that follows from the left side of (\ref{trep}). Thus, 
$\hat{t}^{+}=p^{+}$, $\hat{t}_2$=dilations, etc. In addition to getting the
correct operator products with the currents we will also see that the
dimension of the vertex operator $V_{p^{+}p^{-}}^{r,s}(w)$ is $\Delta
^{r,s}= $ $-r+\frac 1k\left( \frac 14+s^2\right) $.

To define the vertex operator at the quantum level, we begin by preserving
the order of the factors due to the group theoretical origin of their order,
and then apply normal ordering within each factor as follows 
\begin{equation}
V_{p^{+}p^{-}}^{r,s}(z)=e^{iX^{-}(z)p^{+}}<s,p^{+}|\left( :e^{iX_2(z)\hat{t}%
_2}:\right) |s,p^{-}>\left( :e^{ip^{-}\left( q^{+} -\frac
ik\int^zdz^{\prime
}P^{+}(z^{\prime })\,\,:e^{-X_2(z^{\prime })}:\right) }:\right)
\label{vertexop}
\end{equation}
where $X^{+}$ has been written in terms of canonical variables. The leftmost
factor does not need normal ordering (see definition of canonical
variables). This already provides the order of operators. The next step is a
reordering of operators (and picking up factors due to the reordering) for
the purpose of performing computations. A methodical approach for the
computation of operator products is the use of Wick's theorem for free
fields. This requires fully normal ordered expressions. For this purpose we
need to reorder the operators above to rewrite the vertex operator in fully
normal ordered form. This reordering gives the following expression 
\begin{eqnarray}
V_{p^{+}p^{-}}^{r,s}(z) &=&:e^{iX^{-}\left( z\right)p^{+}}
<s,p^{+}|e^{iX_2\left( z\right) \hat{t}_2}  \nonumber  \label{normalV}
e^{ip^{-}q^{+}}
\\
&&\mbox{}\times e^{p^{-}\int^zdz^{\prime }\frac 1k\left( P^{+}(z^{\prime}) 
-\frac{p^{+}}{2z^{\prime }}-\frac{p^{+}}{z-z^{\prime }}\right)
:e^{-X_2(z^{\prime })}:(\frac{z-z^{\prime }}{\sqrt{zz^{\prime }}})^{\frac{2i 
\hat{t}_2}k}}|s,p^{-}>:  \label{orderedV}
\end{eqnarray}
Note that now the whole expression is sandwiched within the normal ordering
columns, and this is what produces the complicated additional factors. In
computations it will be convenient to further rewrite it by expanding the
last exponential as a series and then evaluating the matrix elements at the
end 
\begin{eqnarray}
V_{p^{+}p^{-}}^{r,s}(z) &=&:\exp \left( iX^{-}(z)p^{+}\right) <s,p^{+}|\exp
\left( iX_2(z)\hat{t}_2\right) 
e^{ip^{-}q^{+}} \sum_{n=0}^\infty \frac 1{n!}\left( \frac{%
p^{-}}k\right) ^n  \nonumber \\
&&\mbox{}\times \prod_{i=1}^n\int^zdz_i\left( P^{+}(z_i)-\frac{p^{+}}{2z_i}-%
\frac{p^{+}}{z-z_i}\right) e^{-X_2(z_i)}(\frac{z-z_i}{\sqrt{zz_i}})^
{\frac{2i \hat{t}_2}k}|s,p^{-}>: \\
&=&:Ce^{iX^{-}(z)p^{+}}e^{-\frac12
X_2(z)}e^{ip^{-}q^{+}} \sum_{n=0}^\infty \frac
1{n!}\left( \frac{p^{-}}k\right) ^n  \nonumber \\
&&\mbox{}\times \prod_{i=1}^n\int^zdz_i\left( P^{+}(z_i)-\frac{p^{+}}{2z_i}-%
\frac{p^{+}}{z-z_i}\right) e^{-X_2(z_i)}\left( \frac{z-z_i}{\sqrt{zz_i}}%
\right) ^{-\frac 1k}  \nonumber \\
&&\mbox{}\times J_{2\sigma }\left( 2\sqrt{-p^{+}p^{-}}e^{-\frac
12X_2(z)}\left( \frac{z-z_i}{\sqrt{zz_i}}\right) ^{-\frac 1k}\right) :
\end{eqnarray}
Since the first definition of the vertex operator is not fully normal
ordered, we should show that its vacuum expection value is finite. For this
we use the last form and find 
\begin{eqnarray}
<0|V_{p^{+}p^{-}}^{r,s}(z)|0> &=&:C\sum_{n=0}^\infty \frac 1{n!}\left( \frac{%
p^{-}}k\right) ^n\prod_{i=1}^n\int^zdz_i\left( -\frac{p^{+}}{2z_i}-\frac{%
p^{+}}{z-z_i}\right)  \nonumber \\
&&\mbox{}\,\,\,\,\,\,\times \left( \frac{z-z_i}{\sqrt{zz_i}}\right) ^{-\frac
1k}J_{2\sigma }\left( 2\sqrt{-p^{+}p^{-}}\left( \frac{z-z_i}{\sqrt{zz_i}}%
\right) ^{-\frac 1k}\right) :
\end{eqnarray}
Changing the integration variable to $y_i=2\sqrt{-p^{+}p^{-}}\left( \frac{%
z-z_i}{\sqrt{zz_i}}\right) ^{-\frac 1k}$, we see 
\begin{equation}
<0|V_{p^{+}p^{-}}^{r,s}(z)|0>=C\exp \left( \frac 12 \sqrt{-p^{-}p^{+}}%
\int^\infty dyJ_{2\sigma }(y)\right)
\end{equation}
which is finite.

Later we will be interested in taking the large $k$ limit to show that we
can recover the vertex operator in flat 3D. In this limit our finite normal
ordered vertex operator will be the same as the finite normal ordered vertex
operator in flat Minkowski space-time.

The laborious technical details of the calculation of the operator products
of the currents and stress tensor with the vertex operator are given in the
appendix. With the ordering prescription described above, we find the
correct operator products (\ref{curvertex}). We also show that the operator
product of the vertex operator with the energy-momentum tensor gives the
correct conformal dimension as in (\ref{casim},\ref{weightsrs}) 
\[
\Delta ^{r,s}=\frac{1}{k}\left[ (\hat{t}_{2})^{2}-i\hat{t}_{2}+\hat{t}^{+}%
\hat{t}^{-}\right] =-r+\frac{1}{k}\left( \frac{1}{4}+s^{2}\right) . 
\]

\subsection{Zero curvature limit ($k\rightarrow \infty $) and flat vertex
operator}

\subsubsection{$k\rightarrow \infty $ limit for currents}

If we rescale the affine currents $\alpha _n^\mu \equiv J_n^\mu /\sqrt{k+2}$
and then send $k\rightarrow \infty $ we find 
\[
\begin{array}{ll}
k\neq \infty : & \left[ J_n^\mu ,J_m^\nu \right] =i\varepsilon ^{\mu \nu
\lambda }\eta _{\lambda \rho }J_{n+m}^\rho -(k+2)n\delta _{n+m}\eta ^{\mu
\nu } \\ 
& \left[ \frac{J_n^\mu }{\sqrt{k+2}},\frac{J_m^\nu}{\sqrt{k+2}}\right]
=\frac 1{\sqrt{k+2}}i\varepsilon ^{\mu \nu \lambda }\eta _{\lambda \rho} 
\frac{J_{n+m}^\rho }{\sqrt{k+2}}-n\delta _{n+m}\eta ^{\mu \nu } \\ 
k\rightarrow \infty : & \left[ \alpha _n^\mu ,\alpha _n^\nu \right]
=-n\delta _{n+m}\eta ^{\mu \nu },\quad \eta ^{\mu \nu }=diag\left(
1,-1,-1\right)
\end{array}
\]
Therefore, we get the flat theory in this limit. We do the same with the
free field formulation of the currents, rescale the free fields, and then
identify the flat free fields as the limit of the rescaled free fields.

\begin{eqnarray}
\left( \frac{J^1+J^0}{\sqrt{k+2}}\right) &=&\left( \frac{P^{+}}{\sqrt{ k+2}}%
\right) \\
\left( \frac{J^2}{\sqrt{k+2}}\right) &=&\frac 1{\sqrt{k+2}}:\left( \sqrt{k+2}%
X^{-}\right) \left( \frac{P^{+}}{\sqrt{k+2}}\right) :+\left( \frac{P_2}{%
\sqrt{k+2}}\right) \\
\left( \frac{J^1-J^0}{\sqrt{k+2}}\right) &=&\frac 1{k+2}:\left( \sqrt{ k+2}%
X^{-}\right) \left( \frac{P^{+}}{\sqrt{k+2}}\right) \left( \sqrt{ k+2}%
X^{-}\right) : \\
&&-\frac 2{\sqrt{k+2}}\left( \frac{P_2}{\sqrt{k+2}}\right) \left( \sqrt{ k+2}%
X^{-}\right) +i\partial _z\left( \sqrt{k+2}X^{-}\right)  \nonumber
\end{eqnarray}
As $k\rightarrow \infty $ the expressions in parentheses on both sides are
kept fixed. Therefore, due to the extra factors of $\frac 1{\sqrt{k+2}}$
some terms vanish. Now we identify the flat string coordinates $\breve{X}%
^\mu $ and their conjugate momenta $\breve{P}^\mu $ in the large $k$ limit 
\begin{equation}
\left( \frac{J^\mu }{\sqrt{k+2}}\right) =\breve{P}^\mu
\end{equation}
as the expressions inside parentheses that remain fixed (using $\breve{P}%
^{\pm }=(\breve{P}^1\pm \breve{P}^0)$ and $\breve{X}^{\pm }=\frac 12 (\breve{%
X}^1\pm \breve{X}^0)$ ), then in the large $k$ limit we identify 
\begin{equation}
\begin{array}{c}
\left( \frac{P^{+}}{\sqrt{k+2}}\right) =\breve{P}^{+},\quad \left( \frac{ P_2%
}{\sqrt{k+2}}\right) =\breve{P}^2,\quad i\partial _z\left( \sqrt{k+2}
X^{-}\right) =\breve{P}^{-} \\ 
\left( \sqrt{k+2}X^{+}\right) =\breve{X}^{+},\quad \left( \sqrt{k+2}
X_2\right) =\breve{X}_2,\quad \left( \sqrt{k+2}X^{-}\right) =\breve{X}^{-}.
\end{array}
\end{equation}
The fundamental canonical pairs are $\left( \breve{X}^{-},\breve{P}%
^{+}\right) $ and $\left( \breve{X}^2,\breve{P}^2=i\partial \breve{X}%
^2\right) ,$ which are written in terms of the elementary oscillators
(rescaled ones $\alpha _n^{+,-,2}$). Note that $\breve{X}^{+}$ is derived in
terms of the oscillators in $\breve{P}^{+}$, since $\breve{P}^{+}=i\partial 
\breve{X}^{+}$ in the large $k$ limit according to (\ref{X+}).

In terms of the rescaled variables $\breve{X}^\mu ,\breve{P}^\mu =i\partial
_z\breve{X}^\mu $ , the operator products in the large $k$ limit become
simple for $\breve{X}^{+}$ as well, so that we have the operator products of
the usual flat string fields 
\begin{equation}
k\rightarrow \infty :\quad \breve{X}^{\pm }\left( z\right) \breve{P}^{\mp
}\left( w\right) \sim \breve{X}^2\left( z\right) \breve{P}^2\left( w\right)
\sim \frac i{z-w}
\end{equation}

\subsubsection{Vertex operator in the flat limit}

We want to show that the asymptotic form of the vertex operator as $%
k\rightarrow \infty $ matches the flat form. For this we need to replace $%
p^{+}=\sqrt{k}\breve{p}^{+}$ and $p^{-}=\sqrt{k}\breve{p}^{-}$ so that in
the limit the factors 
\begin{equation}
exp\left( ip^{\pm }X^{\mp }\right) \rightarrow exp\left( i\breve{p}^{\pm }%
\breve{X}^{\mp }\right)
\end{equation}
come out right. Similarly we define $s=\sqrt{k}\breve{p}_2$ so that $\exp
\left( isX_2\right) \rightarrow \exp \left( i\breve{p}_2\breve{X}_2\right) $%
. Replacing the rescaled variables in (\ref{VFrs}) and taking the large $k$
limit we find that the flat vertex operator emerges 
\begin{equation}
V_{\breve{p}^{+}\breve{p}^{-}}^{r,s}(z)\rightarrow e^{i\breve{X}^{-}\breve{p}%
^{+}}e^{i\breve{X}^2\breve{p}^2}e^{i\breve{X}^{+}\breve{p}^{-}}F^{r,s}\left(
k\breve{p}^{+}\breve{p}^{-}\right)
\end{equation}
The overall field independent constant $F^{r,s}\left( k\breve{p}^{+}\breve{p}%
^{-}\right) =\left( k\breve{p}^{+}\breve{p}^{-}\right) ^{i\sqrt{k}\breve{p}%
_2}J_{2m\sqrt{k}}\left( 2\sqrt{-k\breve{p}^{+}\breve{p}^{-}}\right) $, is
given in terms of the Bessel function where we have defined the mass $m$
through $\sigma =\sqrt{k}m$, or 
\begin{equation}
m=\sqrt{r-\breve{p}_2^2}=\sqrt{-\breve{p}^{+}\breve{p}^{-}-\breve{p}_2^2}
\label{mass}
\end{equation}
To evaluate the large $k$ limit of $F^{r,s}\left( k\breve{p}^{+}\breve{p}%
^{-}\right) $ we find it useful to examine the large $k$ limit of the
differential equation in (\ref{Frs}). After rescaling $s=\sqrt{k}\breve{p}_2$
and $\kappa = k\breve{\kappa}$, and taking the large $k$ limit we learn from
the leading term that $F^{r,s}$ behaves like a delta function. After using
the definition of mass given in (\ref{mass}) we obtain 
\begin{equation}
F^{r,s}\rightarrow C\delta \left( \breve{p}^{+}\breve{p}^{-}+\breve{p}%
_2^2+m^2\right)
\end{equation}
where C is a constant. Note that this is the mass shell condition for a flat
string. If we substitute this in the vertex operator, then in the $%
k\rightarrow \infty $ limit we get (up to a constant phase) 
\begin{equation}
\left( V_{p^{+}p^{-}}^{r,s}(z)\right) _{k\rightarrow \infty }=e^{i\breve{X}%
^{-}\breve{p}^{+}}e^{i\breve{X}^2\breve{p}^2}e^{i\breve{X}^{+}\breve{p}%
^{-}}\delta \left( \breve{p}^{+}\breve{p}^{-}+\breve{p}_2^2+m^2\right)
\end{equation}
which is the correct vertex operator in flat 3D-Minkowski space-time, with
the mass shell condition imposed.

\section{AdS-CFT correspondence and unitarity}

\subsection{Boundary operators and principal series}

The results presented in the previous sections establish properties and
vertex operators of the bulk theory in AdS(3). We now discuss the
implications of our results for the proposed structure of the boundary
conformal field theory (CFT) which is claimed to provide a non-perturbative
and second-quantized formulation of string theory on $AdS_{3}\times
S^{3}\times M^{4}$ \cite{GKS}, \cite{ooguri}, \cite{kutseib}. This CFT part
of our discussion is much less complete as compared to our discussion of the
bulk theory, but we feel compelled to include it here because of the
considerable confusion and controversy that exists in the literature.

We are motivated by the proposal of \cite{GKS}, as discussed in more detail
in \cite{ooguri} and as recently presented in \cite{kutseib}. The puzzle we
want to address is how to implement this proposal for constructing the
boundary CFT theory from the bulk string theory without neglecting the
information about the unitarity and the physical state spectrum of the bulk
string theory that we have presented in the previous sections. There is a
puzzle since the vertex operators used in \cite{GKS}\cite{ooguri}\cite
{kutseib} have no relation to the physical states of the theory, and
therefore it is not evident how the bulk and boundary theories are related.
The proposals we make in this section are aimed at trying to close this gap
by connecting our results to those of \cite{GKS}\cite{ooguri}\cite{kutseib}.
This will involve some conceptual issues and some generalizations. In
particular, our proposals for boundary operators include strings that wind
at the boundary. There will still remain unsolved puzzles as will be seen in
the discussion below.

We have particularly emphasized the issues of unitarity and operator-state
correspondence, and therefore we are forced to re-examine the proposal of 
\cite{GKS} in view of the only allowed unitary representation in the bulk
theory discussed above, that is - the principal series representation of the
affine current algebra supplemented by the zero modes. We show that the
proposal of \cite{GKS} has to be generalized and modified to include the
effect of the zero modes, while working in the unitary principal series
representation of the affine current algebra. The vertex operator $%
V_{x^{-},x^{+}}^{r,s}\left( g\right) $ discussed in the previous section of
this paper plays a crucial role in the following construction.

First we briefly summarize the proposal of \cite{GKS} in order to make our
presentation more convenient for the reader. We follow the notation of \cite
{kutseib}. Let the world-sheet currents $k^{a}(z)$ of some world-sheet
current algebra satisfy the following OPEs 
\begin{equation}
k^{a}(z)k^{b}(w)\sim {\frac{1}{2}}{\frac{{k_{G}\delta ^{ab}}}{{(z-w)^{2}}}}+{%
\frac{{f^{abc}k^{c}(w)}}{{(z-w)}}},
\end{equation}
where $k_{G}$ denotes the level of the world-sheet affine Lie algebra $\hat{G%
}$ and $f^{abc}$ are the structure constants of the corresponding Lie group
G. Let the corresponding space-time currents be denoted by $K^{a}(x)$. They
are expected to satisfy the analogous OPEs 
\begin{equation}
K^{a}(x)K^{b}(y)\sim {\frac{1}{2}}{\frac{{k_{G}^{(st)}\delta ^{ab}}}{{%
(x-y)^{2}}}}+{\frac{{f^{abc}K^{c}(y)}}{{(x-y)}}},
\end{equation}
where $k_{G}^{(st)}$ stands for the level of the space-time current algebra.

The proposed form for $K^{a}(x)$ is dictated by symmetry and the fact that $%
K^{a}(x)$ is a $(1,0)$ operator in space-time. Following the proposal of 
\cite{GKS} \cite{kutseib} we present a generalized form of the space-time
current $K^{a}$ while insisting on the only unitary representation and
taking into account the contribution of the zero modes 
\begin{equation}
K^{a}(x)=-{\frac{1}{k}}\int d^{2}z\,\,k^{a}(z)\,\,\bar{J}(x^{+};\bar{z}%
)\,\,\Omega _{x^{-},x^{+}}^{r,s}(z,\bar{z}),  \label{Ka}
\end{equation}
where, the probability distribution $\Omega _{x^{-},x^{+}}^{r,s}(z,\bar{z})$
is defined as in subsection (3.2.1), which we repeat here for convenience 
\begin{equation}
\Omega _{x^{-},x^{+}}^{r,s}\left( z,\bar{z}\right) =\left|
V_{x^{-},x^{+}}^{r,s}\left( g\right) \right| ^{2}.
\end{equation}
Furthermore, $J$ is defined to be, as in \cite{Bars} \cite{kutseib} 
\begin{equation}
J(x^{-};z)\equiv 2x^{-}J^{2}(z)-(J^{1}(z)+J^{0}(z))-\left( x^{-}\right)
^{2}(J^{1}(z)-J^{0}(z)).
\end{equation}
and similarly for $\bar{J}(x^{+};\bar{z}).$ Note that the expression for the
space-time current $K^{a}(x)$ given in \cite{kutseib} formally corresponds
to the form of the probability distribution at $r=0\ $since it then
coincides with the same expression 
\begin{equation}
\Omega _{x^{-},x^{+}}^{0,s}\left( g\right) =\frac{1}{\pi }\left( e^{\phi
/2}\left( x^{+}-\gamma ^{+}\right) \left( x^{-}+\gamma ^{-}\right) +e^{-\phi
/2}\right) ^{-2}  \label{phi1}
\end{equation}
denoted as $\Phi _{1}(x,\bar{x};z,\bar{z})$ in \cite{kutseib}. However, for $%
r\neq 0$ our proposal includes winding strings.

The form of $K^{a}(x)$ may be interpreted as ``dressing'' each worldsheet
operator with a vertex operator that corresponds to a physical state in the
theory. Thus $k^{a}(z)V^{r,s}\left( z,\bar{z}\right) $ and $\bar{J}(x^{+};%
\bar{z})\left( V^{r,s}\left( z,\bar{z}\right) \right) ^{\ast }$ are both
``dressed'', resulting in the factor $\,\Omega _{x^{-},x^{+}}^{r,s}(z,\bar{z}%
)$. Our proposal here differs from \cite{GKS} \cite{kutseib} only by
inserting $\Omega _{x^{-},x^{+}}^{r,s}\left( z,\bar{z}\right) $ which is of
the form $VV^{\ast }$ instead of a single power of the vertex operator $V$.
A single $V$ represents a wavefunction related to a physical state. Since a
wavefunction is generally complex it cannot result in a real $K^{a}(x).$ On
the other hand, the $\Omega _{x^{-},x^{+}}^{r,s}(z,\bar{z})\sim VV^{\ast }$
is a probability density that is $\,$real and carries the relevant
information about the physical state.

We note that we must take the worldsheet conformal dimension $\Delta
^{r,s}=0,$ so that the operators have the correct dimensions. This is
possible provided the zero modes $p^{-},\tilde{p}^{+}$ contribute a non-zero 
$r$. The explicit form of the vertex operator $V_{x^{-},x^{+}}^{r,s}(g\left(
z,\bar{z}\right) )$ and its dimension $\Delta ^{r,s}$ was derived and
discussed in detail in the previous section. In arriving at this expression,
in our case we only used the principal series representation at $r=0$, and
we never had to refer to the non-existent discrete series representation
that is not consistent with unitarity. This approach maintains
operator-state correspondence as well as unitarity, since we have used only
the positive norm physical states that correspond to the probability $\Omega
_{x^{-},x^{+}}^{r,s}\left( g\right) $. In the next section we will briefly
discuss the OPEs of the space-time currents $K^{a}$ and compare to the
results of \cite{kutseib}.

We can easily extend our discussion and consider the generators of the
space-time Virasoro symmetry. The form of the space-time stress tensor $%
T^{st}(x)$ is also dictated by symmetry and the fact that $T^{st}(x)$ should
be a $(2,0)$ operator in space-time. Again, we generalize the proposal
presented in \cite{GKS} \cite{kutseib}, and present formulae valid for the
unitary principal series representation of the affine current algebra
supplemented with zero modes 
\begin{equation}
T^{st}(x)={\frac{1}{2k}}\int d^{2}z\left( \partial _{x^{-}}J({x}^{-};{z}%
)\partial _{x^{-}}\Omega _{x^{-},x^{+}}^{r,s}\left( g\right) +2\partial
_{x^{-}}^{2}J({x}^{-};{z})\Omega _{x^{-},x^{+}}^{r,s}\left( g\right) \right) 
\tilde{J}(x^{+};\bar{z}).  \label{cftT}
\end{equation}
As before, the expression given in \cite{kutseib} is formally the special
case of this formula at $r=0$, however in our case we do not refer to the
discrete series representation which is not consistent with unitarity. In
our version unitarity and operator-state correspondence is preserved by
using only the probability $\Omega _{x^{-},x^{+}}^{r,s}\left( g\right) $
associated with physical states. The corresponding OPE 
\begin{equation}
T^{st}(x)T^{st}(y)\sim {\frac{1}{2}}{\frac{c^{st}}{{(x-y)^{4}}}}+{\frac{%
2T^{st}(y)}{{(x-y)^{2}}}}+{\frac{\partial _{y}T^{st}(y)}{{(x-y)}}},
\end{equation}
in principle determines the value of the central charge of the boundary
conformal field theory $c^{st}$. This value should be compared to the
classical expression for the central charge of the Virasoro algebra at the
boundary of $AdS_{3}$ found in the framework of pure three-dimensional
quantum gravity \cite{bh}. In the next section we will briefly discuss the
OPEs of the space-time stress tensor $T(x)$ and compare to the results of 
\cite{kutseib}.

Note that besides coinciding with the expressions in \cite{kutseib} at $r=0,$
our proposal takes into account the winding strings when $r\neq 0.$

\subsection{Operator products}

In this section we want to discuss the OPEs of the space-time currents $%
K^{a}(x)$ in the space-time current algebra and the OPEs of the space-time
stress tensor $T^{st}(x)$ in the space-time Virasoro algebra, given the
general expressions for $K^{a}(x)$ and $T^{st}(x)$ presented above.

In principle, the relevant OPEs should be evaluated for all values of the
field $\phi \left( z,\bar{z}\right) $. To accomplish this it seems necessary
to use the general OPEs of two vertex operators $V_{x^{-},x^{+}}^{r,s}(g)$.
These expressions can be evaluated in principle by using the free field
representation discussed in this paper. The corresponding analytically
continued expression (relevant for the $SL(2,C)/SU(2)$ WZW model) have been
presented by Teschner \cite{teschner}, \cite{teschner2}, however only at $%
r=0 $. The general form of the $%
V_{x^{-},x^{+}}^{r,s}(g)V_{x^{-},x^{+}}^{r,s}(g)$ turns out to be rather
complicated, even at $r=0$, as in can be seen from \cite{teschner}, \cite
{teschner2}. However, provided one takes the limit $\phi \rightarrow \infty $%
, which corresponds to the boundary of $AdS_{3}$, the corresponding
semiclassical expression for the OPEs of two vertex operators $%
V_{x^{-},x^{+}}^{r,s}(g)$ simplify considerably.

In particular, we have already seen that for $\phi \rightarrow \infty $ 
\begin{equation}
\Omega _{x^{-},x^{+}}^{r,s}(g)\rightarrow \delta (x^{+}-\gamma ^{+})\delta
(x^{-}-\gamma ^{-})\left( \left[ c_{1}\log ((x^{+}-\gamma ^{+})(x^{-}-\gamma
^{-}))+c_{1}\phi +c_{2}\right] \right) ^{2},
\end{equation}
where $c_{1}=0,\,c_{2}=1$ if $r=0$, and $c_{1}\phi +c_{2}$ is dropped if $%
r\neq 0$. When $r=0$ only the delta function survives in this limit. From
this expression it follows that $\Omega _{j_{1}}\equiv \Omega
_{x_{1}^{-},x_{1}^{+}}^{0,s_{1}}(g)$ and $V_{j_{2}}\equiv
V_{x_{2}^{-},x_{2}^{+}}^{0,s_{2}}(g)$ satisfy (with $j_{1,2}=-1/2+is_{1,2}$,
as implied by the unitary principal series representation) 
\begin{eqnarray}
\lim_{z_{1}\rightarrow z_{2}}\Omega _{j_{1}}(x_{1},\bar{x_{1}};z_{1},\bar{%
z_{1}})\Omega _{j_{2}}(x_{2},\bar{x_{2}};z_{2},\bar{z_{2}}) &=&\delta
(x_{1}^{+}-x_{2}^{+})\delta (x_{1}^{-}-x_{2}^{-})\,\Omega
_{j_{1}+j_{2}-1}(x_{2},\bar{x_{2}};z_{2},\bar{z_{2}}) \\
\lim_{z_{1}\rightarrow z_{2}}\Omega _{j_{1}}(x_{1},\bar{x_{1}};z_{1},\bar{%
z_{1}})V_{j_{2}}(x_{2},\bar{x_{2}};z_{2},\bar{z_{2}}) &=&\delta
(x_{1}^{+}-x_{2}^{+})\delta (x_{1}^{-}-x_{2}^{-})\,V_{j_{1}+j_{2}-1}(x_{2},%
\bar{x_{2}};z_{2},\bar{z_{2})} \\
\lim_{z_{1}\rightarrow z_{2}}V_{j_{1}}(x_{1},\bar{x_{1}};z_{1},\bar{z_{1}}%
)V_{j_{2}}(x_{2},\bar{x_{2}};z_{2},\bar{z_{2}}) &=&\sqrt{\pi }\delta \left( 
\sqrt{\left( x_{1}^{+}-x_{2}^{+}\right) \left( x_{1}^{-}-x_{2}^{-}\right) }%
\right) V_{j_{1}+j_{2}-\frac{1}{2}}(x_{2},\bar{x_{2}};z_{2},\bar{z_{2})} 
\nonumber
\end{eqnarray}
For the last case we used (with $\varepsilon =\exp \left( -\phi /2\right)
\rightarrow 0$) 
\begin{equation}
\frac{1}{\varepsilon ^{-1}u^{2}+\varepsilon }=\frac{1}{2i}\left( \frac{1}{%
u-i\varepsilon }-\frac{1}{u+i\varepsilon }\right) =\pi \delta \left(
u\right) ,
\end{equation}
which is different than \cite{kutseib}. Note that these expressions are
obviously consistent with the fact that we are working in the principal
series representation $j=-1/2+is$ which closes under operator products.
These expressions can be used as a semiclassical limit of the complicated
OPE involving probability distributions $\Omega _{x^{-},x^{+}}^{0,s}(g)$ or
vertex operators $V_{x^{-},x^{+}}^{0,s}(g)$ at $r=0$. Notice that the
analogous expressions used in \cite{kutseib} (eqs. (2.34) and (2.35) of \cite
{kutseib}) are not consistent with the closure in the principal series
representation.

Nevertheless, we use the semiclassical reasoning of \cite{kutseib}, to argue
that the space-time currents $K^{a}(x)$ and the space-time stress tensor $%
T^{st}(x)$ defined above satisfy the OPEs expected from the space-time
current algebra and the Virasoro algebra respectively. However, it should be
noted that the $\phi \rightarrow \infty $ asymptotics of the general vertex
operator $V_{x^{-},x^{+}}^{r,s}(g)$ involves, apart from the delta function,
a log piece as well, so that the general formula for the semiclassical limit
of the OPE of two vertex operators is not as simple as the $r=0$ case, and
the semiclassical analysis along the lines of \cite{kutseib} becomes more
involved. These refinements are still to be worked out (in principle, by
using our free fields), but it is likely that the exact formulas at $r\neq 0$
and general $\phi $ will be similar.

With the caviats above in mind, following the same steps presented in \cite
{kutseib} one sees that 
\begin{equation}
K^{a}(x)K^{b}(y)\sim {\frac{1}{2}}{\frac{{Ik_{G}^{(st)}\delta ^{ab}}}{{%
(x-y)^{2}}}}+{\frac{{f^{abc}K^{c}(y)}}{{(x-y)}}},
\end{equation}
where in our case the operator $I$ (in the notation of \cite{kutseib}) is
defined as 
\begin{equation}
I\equiv {\frac{1}{k^{2}}}\int d^{2}zJ({x};{z})\bar{J}(\bar{x};\bar{z})\Omega
_{x^{-},x^{+}}^{r,s}(g).
\end{equation}
The analogous formula given in \cite{kutseib} is formally once again a
special case of this expression for $r=0$. However, we emphasize that in
general (i.e. when $r\neq 0$) it is much harder to argue (as done in \cite
{kutseib}) that the value of the expectation value of the operator $I$
corresponds precisely to the value of the central charge expected from the
classical analysis of \cite{bh}. Similar observation are valid for the OPEs
of the space-time stress tensor $T^{st}(x)$.

We conclude this section with the following remark: In many papers on $%
AdS_{3}/CFT_{2}$ correspondence \cite{GKS} \cite{ooguri} \cite{kutseib} it
is claimed that the Casimir label $j$ of the bulk $SL(2,R)$ current algebra
is related to the spacetime Virasoro highest weight $h$ (because the global $%
SL(2,R)$ symmetry of the bulk $AdS_{3}$ corresponds to the global conformal
symmetry of the boundary $CFT_{2}$ generated by $L_{0}$ and $L_{\pm }$ \cite
{MS}), and since $h$ is a conformal weight of the boundary $CFT_{2}$, $j$
has to be real, forcing one to consider the non-unitary discrete series
representation. We comment on this argument as follows.

The Virasoro heighest weight $h$ is undoubtedly real since it is associated
with the hermitian spacetime conformal operator $T^{st}\left( x\right) $.
However, the relation between $h$ and the complex $j$ is not necessarily a
simple one. The spacetime $h$ corresponds to an operator that forms a
representation of the space-time theory $K^{a},T^{st}$, and such an operator
is constructed from bulk vertex operators $V^{r,s}$ that carry the label $j,$
or more precisely $s,r$. As we have seen in the previous section, our
construction differs from previous ones and the computation of $h$ is
thereby altered. For example, the probability density $\Omega
_{x^{-},x^{+}}^{r,s}(g)$ at $r=0$ has the space-time weight $h=-(j+j^{\ast
})=1$ which is what was needed to construct the $\Phi _{1}(x,\bar{x};z,\bar{z%
})$ in \cite{kutseib} at $r=0,$ yet the value of $j$ is complex.

This example serves to illustrate that one should distinguish the
representation space of physical states in the space-time theory from the
representation space of physical states in the bulk theory, although one
expects some relationship between them, and therefore that $j$ in the bulk
theory is quite happy to be complex while $h$ in the boundary theory is
real. As we have insisted in this paper, the bulk theory is unitary with its
physical states appearing in the principal series representation of the
affine currents $j^{\mu }\left( z\right) ,$ with $j=-1/2+is$, and distorted
by the effects of zero modes that describe winding strings, represented by
sectors labelled with a quantum number $r$. The dependence of $h$ on $s,r$
(or $j)$ for the representation space of the space-time theory is currently
an unsolved problem. Likewise the relation of the physical states of the
bulk to the physical states of the boundary theory remains to be clarified.

In this paper we presented what we believe to be the correct description of
the Hilbert space of the bulk theory, but our work remains incomplete with
regards to the Hilbert space of the boundary theory, and its relationship to
the bulk vertex operators. Thus we think that the right interpretation of $j$
consistent with the unitary principal series representation, supplemented
with zero modes, is the one already given in the second part of this paper.
The ``problems'' associated with alleged non-unitarity and failure of
operator-state correspondence for the case of string theory on $AdS_{3}$, as
discussed in many papers in the literature, simply do not exist in our
construction. To fully establish the AdS$_{3}$-CFT conjecture, the physical
vertex operators need to be used as building blocks, as we have shown
partially in this paper, and the refinements discussed in this paper for the
vertex operator need to be implemented more rigorously.

\section{Conclusions}

To summarize, in this paper we have discussed string theory on $%
AdS_{3}\times S^{3}\times M^{4}$ while emphasizing the issues of unitarity
and state-operator correspondence. In particular we have re-examined the $%
AdS_{3}$-$CFT_{2}$ correspondence in the Minkowski signature, by taking into
account the only allowed unitary representation, the principal series of
SL(2,R) supplemented with the zero modes. The zero modes play an important
role in the description of on-shell physical states or vertex operators.
Without them a unitary formulation of the on-shell theory is not possible.
Also, the zero modes describe the winding of long strings around the $%
AdS_{3} $ boundary.

The theory is presented as part of the supersymmetric WZW model that
includes the supergroup SU$\left( 2/1,1\right) $ (or OSp$\left( 4/2\right) $
or $D\left( 2,1;\alpha \right) $) with central extension $k$. A free field
representation is given and the vertex operators are constructed in terms of
free fields in SL(2,R) principal series representation bases that are
labeled by position space or momentum space at the boundary of $AdS_{3}$.
Our vertex operator includes factors that have been missed in the recent
literature on the $AdS_{3}/CFT_{2}$ correspondence. These factors are
related to winding strings. We have shown explicitly that our vertex
operator has the correct operator products with the currents and stress
tensor, all of which are constructed from free fields. We have also shown
that in the limit when $AdS_{3}$ tends to flat 3D-Minkowski space ( $%
k\rightarrow \infty $), the AdS$_{3}$ vertex operators in momentum space
tend to the vertex operators of flat 3D-string theory (and furthermore the
theory readjusts smoothly in the rest of the dimensions in the same limit).

There remains to compute in strictly string theoretical language the various
correlation functions or operator products that would verify or refine the
AdS-CFT correspondence. In principle our free field vertex operators can be
used to compute any correlation function. We intend to discuss these
computations in the future.

{\large {\bf Acknowledgements}}

We thank T. Eguchi, K. Ito, Y. Sugawara, S.K. Yang, H. Ooguri and D.
Kutasov for discussions. Some of the progress reported in this paper was
done during a collaborative research visit to Japan supported partially by
the NSF and JSPS. We thank our Japanese collaborators for their hospitality.

\appendix

\section{Appendix: Operator products}

In this appendix we are going to describe in detail the full quantum
treatment of the operator products of the vertex operator with currents and
energy-momentum tensor. This calculation is done using the same notation as
in \cite{Bars} which we keep here so that we can use previous results.
Therefore one should be careful about the following relations between the
notation in this appendix and the notation in the previous sections of this
paper: 
\begin{equation}
\begin{array}{rrrr}
{\rm {Previous\,\,sections:}} & X^{-}(z), & X_2(z), & X^{+}(z) \\ 
{\rm {Appendix:}} & -X^{-}(z), & -\frac 2k u(z), & -X^{+}(z)
\end{array}
\label{a1}
\end{equation}

In the first subsection of the appendix we are going to describe the quantum
ordered expression for the 2$\times 2$ group element ($j=1/2$
representation) and discuss its operator product with the energy-momentum
tensor. Besides serving to double check the result that follows from
products with currents $J^\mu \left( z\right) g\left( \omega \right) \sim
t^\mu g\left( \omega \right) /\left( z-w\right) $ \cite{Bars}, this
calculation is also useful to clarify some delicate points of the
calculation in more general representations $j$. In the second subsection we
are going to discuss operator products of the full vertex operator with the
currents and also find the conformal dimension of the vertex operator as
stated in (\ref{weightsrs}).

The method of computation is based on Wick's theorem for free fields. The
expressions for contractions between free fields which will be used in the
following computations are as follows 
\begin{eqnarray}
&<X^{-}(z)\,P^{+}(w)>\equiv \frac i{2w}+\frac i{z-w},\quad
<u(z)\,S(w)>\equiv \frac k2\left( \frac i{2w}+\frac i{z-w}\right) \\
&<P^{+}(z)\,X^{-}(w)>\equiv \frac i{2z}+\frac{-i}{z-w},\quad
<u(z)\,u(w)>\equiv -\frac k2\ln \left( \frac{z-w}{\sqrt{zw}}\right) .
\end{eqnarray}
The terms of the form $\frac i{2z}$ arise from the careful ordering of zero
modes such that hermiticity is respected at every step \cite{Bars}. The
definition of the contraction includes this effect of the zero modes.

\subsection{Operator products with the group element}

We have the quantum operator version of the left moving group element $%
g_{mn} $ , as in (\ref{L}) modulo the map (\ref{a1}) 
\begin{eqnarray}
g_{11} &=&:e^{-\frac{u(w)}k}:,\quad g_{12}=:e^{\frac{-u\left( w\right) }%
k}:X^{+}\left( w\right) \\
g_{21} &=&X^{-}(w):e^{-\frac{u(w)}k}:,\quad g_{22}=e^{\frac{u(w)}k}+e^{\frac{%
-u(w)}k}X^{-}X^{+}
\end{eqnarray}
The orders of the operators are important in these expressions. This order
of operators is dictated by the orders of the matrices that make up $g.$
This order will also be respected in more general representations. Then
these $g_{mn}$ are not fully normal ordered (recall that the dependent
operator $X^{+}$ given in (\ref{X+}) contains $P^{+}$ and $:e^{2u(w^{\prime
})/k}:$ ). In order to do the calculations of operator products by using
Wick's theorem, first we should rewrite each $g_{mn}$ in normal ordered
form, by keeping all the terms that arise from re-ordering the operators, as
follows

\begin{eqnarray}
g_{11} &=&:e^{-\frac{u(w)}k}:  \label{orderedg} \\
g_{12} &=&-i\int^wdw^{\prime }\frac{P^{+}(w^{\prime })}k\left( :e^{\frac{%
2u(w^{\prime })}k-\frac{u(w)}k}:\right) (\frac{w-w^{\prime }}{\sqrt{%
ww^{\prime }}})^{\frac 1k} \\
g_{21} &=&X^{-}(w):e^{-\frac{u(w)}k}: \\
g_{22} &=&-i\int^wdw^{\prime }\frac 1k\left( :X^{-}(w)P^{+}(w^{\prime })e^{%
\frac{2u(w^{\prime })}k-\frac{u(w)}k}:\right) (\frac{w-w^{\prime }}{\sqrt{%
ww^{\prime }}})^{\frac 1k} \\
&&+\int^wdw^{\prime }\frac 1k\left( \frac 1{2w^{\prime }}+\frac
1{w-w^{\prime }}\right) (\frac{w-w^{\prime }}{\sqrt{ww^{\prime }}})^{\frac
1k}\left( :e^{\frac{2u(w^{\prime })}k-\frac{u(w)}k}:\right) \\
&&+\left( :e^{\frac{2u(w)}k}:\right) \left( :e^{\frac{-u(w)}k}:\right)
\end{eqnarray}
In these expressions the ordering of the zero modes of $:e^{\gamma u(w)/k}:$%
, where $\gamma $ is a constant, is defined as 
\begin{equation}
\left[ :e^{\gamma u(w)/k}:\right] _{zero\,modes}=e^{-i\gamma s\ln
z/k}\,\,e^{\gamma u_0/k}\,\,z^{(\gamma ^2-2\gamma )/4k}.
\end{equation}

Operator products of the currents with the group element were already
presented in \cite{Bars}. The operator product of the energy-momentum tensor 
\begin{eqnarray}
T(z) &=&:P^{+}i\partial X^{-}:+T_{P_2}(z)  \label{enmom} \\
T_{P_2}(z) &=&\frac 1k\left( :{P_2}^2:-\frac iz\partial (z{P_2})+\frac
1{4z^2}\right)
\end{eqnarray}
with the group element is given as

\begin{eqnarray}
T(z)\times g(w) &\rightarrow &\frac{hg\left( w\right) }{\left( z-w\right) ^2}%
+\frac{\partial _wg\left( w\right) }{\left( z-w\right) },\quad \\
h &=&-\frac{j\left( j+1\right) }k=-\frac{3/4}k
\end{eqnarray}

In order to calculate this expression firstly we need the result of the
following two operator products: the operator products of $T(z)$ with $%
:e^{-u(w)/k}:$ and $:e^{2u(w^{\prime })/k}:$ 
\begin{eqnarray}
T(z)\times &:&e^{-u(w)/k}:=\frac{-3/4k}{(z-w)^2}:e^{-u(w)/k}:+\frac
1{z-w}:\partial _we^{-u(w)/k}:  \label{temu} \\
T(z)\times &:&e^{2u(w^{\prime })/k}:=\frac 1{z-w^{\prime }}:\partial
_{w^{\prime }}e^{2u(w^{\prime })/k}:
\end{eqnarray}
In this calculations one should be careful about the normal ordering of the
zero modes as described above. Using (\ref{temu}), we find 
\begin{eqnarray}
T\times g_{11} &\rightarrow &\frac{-3/4k}{(z-w)^2}g_{11}+\frac
1{(z-w)}\partial _wg_{11} \\
T\times g_{21} &\rightarrow &:X^{-}<T_{P_2}e^{-u(w)/k}>:+:i\partial
_zX^{-}<P^{+}X^{-}>e^{-u(w)/k}>: \\
&=&\frac{-3/4k}{(z-w)^2}g_{21}+\frac 1{(z-w)}\partial _wg_{21}
\end{eqnarray}
The operator product of $T(z)$ with $g_{12}$ is expected to be 
\begin{equation}
T\times g_{12}\rightarrow \left( -\frac 3{4k}\right) \frac{g_{12}\left(
w\right) }{(z-w)^2}+\frac{\partial _wg_{12}}{(z-w)}  \label{Tg1212}
\end{equation}
where 
\begin{eqnarray}
i\partial _wg_{12} &=&\int^wdw^{\prime }\frac{P^{+}(w^{\prime })}k\partial
_w\left( :e^{\frac{2u(w^{\prime })}k-\frac{u(w)}k}:\right) (\frac{%
w-w^{\prime }}{\sqrt{ww^{\prime }}})^{\frac 1k}  \label{Tg121} \\
&&+\frac{P^{+}(w)}k\left( :e^{\frac{2u(w)}k}:\right) \left( :e^{-\frac{u(w)}%
k}:\right)  \label{Tg122} \\
&&+\int^wdw^{\prime }\frac{P^{+}(w^{\prime })}k\left( :e^{\frac{2u(w^{\prime
})}k-\frac{u(w)}k}:\right) \frac{1/k}{w-w^{\prime }}(\frac{w-w^{\prime }}{%
\sqrt{ww^{\prime }}})^{\frac 1k}  \label{Tg123} \\
&&-\int^wdw^{\prime }\frac{P^{+}(w^{\prime })}k\left( :e^{\frac{2u(w^{\prime
})}k-\frac{u(w)}k}:\right) \frac{1/2k}w(\frac{w-w^{\prime }}{\sqrt{%
ww^{\prime }}})^{\frac 1k}  \label{Tg124}
\end{eqnarray}
To see how this arises through operator products we list the contractions to
be calculated 
\begin{eqnarray}
T\times g_{12} &\rightarrow &:iP^{+}\left( -i\int^wdw^{\prime }\frac
1k<\partial _zX^{-}P^{+}(w^{\prime })>:e^{\frac{2u(w^{\prime })}k-\frac{u(w)}%
k}:(\frac{w-w^{\prime }}{\sqrt{ww^{\prime }}})^{\frac 1k}\right)
\label{Tg12con1} \\
&&-i\int^wdw^{\prime }\frac{P^{+}(w^{\prime })}k\left( :<T_{P_2}e^{\frac{%
2u(w^{\prime })}k}>e^{-\frac{u(w)}k}:\right) (\frac{w-w^{\prime }}{\sqrt{%
ww^{\prime }}})^{\frac 1k}  \label{Tg12con2} \\
&&-i\int^wdw^{\prime }\frac{P^{+}(w^{\prime })}k\left( :e^{\frac{%
2u(w^{\prime })}k}<T_{P_2}e^{-\frac{u(w)}k}>:\right) (\frac{w-w^{\prime }}{%
\sqrt{ww^{\prime }}})^{\frac 1k}  \label{Tg12con3} \\
&&-\frac{2i}k\int^wdw^{\prime }\frac{P^{+}(w^{\prime })}k\left( :<{P_2}e^{%
\frac{2u(w^{\prime })}k}><{P_2}e^{-\frac{u(w)}k}>:\right) (\frac{w-w^{\prime
}}{\sqrt{ww^{\prime }}})^{\frac 1k}  \label{Tg12con4}
\end{eqnarray}
where the factor of $2$ in the last term comes from the permutations of ${P_2%
}$'s. The first term in (\ref{Tg1212}) and the term in (\ref{Tg121}) are
obtained from contraction in (\ref{Tg12con3}). The contraction in (\ref
{Tg12con4}) is equal to 
\begin{eqnarray}
&&-\frac ik\int^wdw^{\prime }\frac{P^{+}(w^{\prime })}k\frac
1{(z-w)(z-w^{\prime })}\left( :e^{\frac{2u(w^{\prime })}k-\frac{u(w)}%
k}:\right) (\frac{w-w^{\prime }}{\sqrt{ww^{\prime }}})^{\frac 1k}
\label{Tg12dc} \\
&&+\frac ik\int^wdw^{\prime }\frac{P^{+}(w^{\prime })}k\frac
1{2z(z-w)}\left( :e^{\frac{2u(w^{\prime })}k-\frac{u(w)}k}:\right) (\frac{%
w-w^{\prime }}{\sqrt{ww^{\prime }}})^{\frac 1k}  \label{Tg12dc2} \\
&&+\frac ik\int^wdw^{\prime }\frac{P^{+}(w^{\prime })}k\frac
1{2z(z-w^{\prime })}\left( :e^{\frac{2u(w^{\prime })}k-\frac{u(w)}k}:\right)
(\frac{w-w^{\prime }}{\sqrt{ww^{\prime }}})^{\frac 1k}  \label{Tg12dc3}
\end{eqnarray}
The term (\ref{Tg12dc2}) is just the term in (\ref{Tg124}). Writing 
\begin{equation}
\frac 1{(z-w^{\prime })}=\frac 1{(w-w^{\prime })}-(z-w)\frac 1{(z-w^{\prime
})}\frac 1{(w-w^{\prime })}  \label{relation}
\end{equation}
in the first term (\ref{Tg12dc}) one gets 
\begin{eqnarray}
&&-\frac i{z-w}\int^wdw^{\prime }\frac{P^{+}(w^{\prime })}k\left( :e^{\frac{%
2u(w^{\prime })}k-\frac{u(w)}k}:\right) \frac{1/k}{w-w^{\prime }}(\frac{%
w-w^{\prime }}{\sqrt{ww^{\prime }}})^{\frac 1k}  \label{partial} \\
&&-i\int^wdw^{\prime }\frac{P^{+}(w^{\prime })}k\frac 1{(z-w^{\prime
})}\left( :e^{\frac{2u(w^{\prime })}k-\frac{u(w)}k}:\right) \partial
_{w^{\prime }}(w-w^{\prime })^{\frac 1k}(\frac 1{\sqrt{ww^{\prime }}%
})^{\frac 1k}  \label{partia}
\end{eqnarray}
where the first term is equal to the term in (\ref{Tg123}). The partial
integration of the second term (\ref{partia}) gives 
\begin{eqnarray}
&&-\frac i{z-w}\frac{P^{+}(w)}k\left( :e^{\frac{2u(w)}k}:\right) \times
\left( :e^{-\frac{u(w)}k}:\right)  \label{partial21} \\
&&+i\int^wdw^{\prime }\frac{\partial _{w^{\prime }}P^{+}(w^{\prime })}k\frac
1{(z-w^{\prime })}\left( :e^{\frac{2u(w^{\prime })}k-\frac{u(w)}k}:\right) (%
\frac{w-w^{\prime }}{\sqrt{ww^{\prime }}})^{\frac 1k}  \label{partial22} \\
&&+i\int^wdw^{\prime }\frac{P^{+}(w^{\prime })}k\frac 1{(z-w^{\prime
})^2}\left( :e^{\frac{2u(w^{\prime })}k-\frac{u(w)}k}:\right) (\frac{%
w-w^{\prime }}{\sqrt{ww^{\prime }}})^{\frac 1k}  \label{partial23} \\
&&+i\int^wdw^{\prime }\frac{P^{+}(w^{\prime })}k\frac 1{(z-w^{\prime
})}\left( :\partial _{w^{\prime }}e^{\frac{2u(w^{\prime })}k-\frac{u(w)}%
k}:\right) (\frac{w-w^{\prime }}{\sqrt{ww^{\prime }}})^{\frac 1k}
\label{partial24} \\
&&-i\int^wdw^{\prime }\frac{P^{+}(w^{\prime })}k\frac 1{(z-w^{\prime
})}\left( :e^{\frac{2u(w^{\prime })}k-\frac{u(w)}k}:\right) \frac{1/k}{%
2w^{\prime }}(\frac{w-w^{\prime }}{\sqrt{ww^{\prime }}})^{\frac 1k}
\label{partial25}
\end{eqnarray}
In this expression the first term (\ref{partial21}) gives the term in (\ref
{Tg122}), the terms (\ref{partial22},\ref{partial23}) combined cancel the
result of the term in (\ref{Tg12con1}), likewise the fourth term (\ref
{partial24}) cancels the result of the term in (\ref{Tg12con2}). Finally the
last term (\ref{partial25}) gives a finite result combined with the term in (%
\ref{Tg12dc3}).

The calculation of the operator product $T(z)$ $\times g_{22}(w)$ is similar
to the above calculation. Since in the integrand there are extra $w$
dependent functions, one will see their $w$-derivatives appear in the
operator product expansion. The double contraction $:i<P^{+}(z)X^{-}(w)><%
\partial _zX^{-}(z)P^{+}(w^{\prime })>:$ gives the term coming from the $w$%
-derivative of $1/(w-w^{\prime })$. Other than this, the contraction $%
:<P^{+}(z)X^{-}(w)>i\partial _zX^{-}(z)P^{+}(w^{\prime }):$ gives the term
that contains $\partial _wX^{-}(w)$.

If one calculates the conformal weight of $:e^{\frac{u(w)}k}:$ one finds
that it is equal to $1/\left( 4k\right) $, but not $-3/\left( 4k\right) $.
However, one should be careful about a special term coming from the double
contraction of $:{P_2}^2:$ with $g_{22}$. Following the same steps as in the
previous calculation we found a term similar to the term in (\ref{partia}): 
\begin{eqnarray}
&&-i\int^w\frac{dw^{\prime }}k\left( X^{-}(w)P^{+}(w^{\prime })+\frac
i{2w^{\prime }}+\frac i{w-w^{\prime }}\right) \left( :e^{\frac{2u(w^{\prime
})}k-\frac{u(w)}k}:\right) \frac 1{(z-w^{\prime })} \\
&&\hspace{0.5in}\times \partial _{w^{\prime }}(w-w^{\prime })^{\frac
1k}(\frac 1{\sqrt{ww^{\prime }}})^{\frac 1k}
\end{eqnarray}
Here the $1/(w-w^{\prime })$ part is new compared to the previous
calculation. The other parts work as before. The partial integration of this
extra term contains in particular 
\begin{eqnarray}
&&-\frac ik\left[ \left( \frac i{w-w^{\prime }}\right) \frac 1{(z-w^{\prime
})}:e^{\frac{2u(w^{\prime })}k-\frac{u(w)}k}:(\frac{w-w^{\prime }}{\sqrt{%
ww^{\prime }}})^{\frac 1k}\right] _{w^{\prime }=w}  \label{extra1} \\
&=&-\frac ik\left[ \left( \frac i{w-w^{\prime }}\right) \frac 1{(z-w^{\prime
})}:e^{\frac{2u(w^{\prime })}k}:\times :e^{-\frac{u(w)}k}:\right]
_{w^{\prime }=w}  \label{extra2} \\
\quad &=&-\frac{i/k}{z-w}\left[ \left( \frac i{w-w^{\prime }}\right) :e^{%
\frac{2u(w^{\prime })}k}:\times :e^{-\frac{u(w)}k}:\right] _{w^{\prime }=w}
\label{extra3} \\
&&-\frac{1/k}{(z-w)^2}\left( :e^{\frac{2u(w)}k}:\right) \times \left( :e^{-%
\frac{u(w)}k}:\right)  \label{extra4}
\end{eqnarray}
The term (\ref{extra3}) comes from the derivative of $g_{22}(w)$. The term (%
\ref{extra4}) is the term that we were looking for to obtain the correct
conformal weight $-3/\left( 4k\right) $. Changing $:e^{\frac{u(w)}k}:$ to $%
:e^{\frac{2u(w)}k}:\times :e^{-\frac{u(w)}k}:$ does not change its conformal
dimension. Adding its conformal dimension, $1/4k$, to the coefficient of the
term in (\ref{extra4}) we again find $-3/4k$. Therefore the operator product
of $g_{22}(w)$ with $T(z)$ dictates that the quantum theory be written in
the normal ordered form form $:e^{\frac{2u(w)}k}:\times :e^{-\frac{u(w)}k}:$
instead of $:e^{\frac{u(w)}k}:$.

This calculation shows that the conformal dimension of each entry in the
group element $g_{mn}$ is $-3/4k$.

\subsection{Operator products with the vertex operator}

In order to simplify the expressions of the operator products we are going
to use the expression (\ref{orderedV}) for the vertex operator, which we
write again for convenience in the form 
\begin{equation}
V_{p^{+}p^{-}}^{r,s}(w)=:f(w)<jp^{+}|g(w)h(w)|jp^{-}>:
\end{equation}
where 
\begin{eqnarray}
f(w) &=&e^{-iX^{-}(w)p^{+}} \\
g(w) &=&:e^{-i\frac{2u(w)}kt_2}: \\
h(w) &=&:\left( e^{-p^{-}\int^w\frac{dw^{\prime }}k\left( P^{+}(w^{\prime })+%
\frac{p^{+}}{2w^{\prime }}+\frac{p^{+}}{w-w^{\prime }}\right) :e^{\frac{%
2u(w^{\prime })}k}:\left( \frac{w-w^{\prime }}{\sqrt{ww^{\prime }}}\right) ^{%
\frac{2it_2}k}}\right) :
\end{eqnarray}
In the following expressions we will often have $\partial _{p^{+}}g(w)$.
Even though in the above expression $g(w)$ seem not to contain $p^{+}$
explicitly, one sees $p^{+}$ dependence after calculating its matrix element
between the states $<s,p^{+}|$ and $|s,p^{-}>$ (\ref{braket}). Keeping this
in mind, we will also often omit the states $<s,p^{+}|$ and $|s,p^{-}>$ in
most of the following expressions for simplicity, but their presence is
implied.

In the case of operator product of $J^{+}$ with the vertex operator there is
only one contraction 
\begin{eqnarray}
J^{+}(z)\times V_{p^{+}p^{-}}^{r,s}(w) &\rightarrow &:<P^{+}f>gh:\,=\,-\frac{%
p^{+}}{z-w}V_{p^{+}p^{-}}^{r,s}(w) \\
&=&-\frac 1{z-w}<s,p^{+}|t^{+}V_{\hat{\jmath}}(w)|s,p^{-}>
\end{eqnarray}
For the operator product of $J_2$ with the vertex operator we need to
calculate the following contractions 
\begin{eqnarray}
J_2(z)\times V_{p^{+}p^{-}}^{r,s}(w) &\rightarrow
&:X^{-}<P^{+}f>gh:+:<P^{+}f>g<X^{-}h>:  \nonumber \\
&&+:f<{P_2}g>h:+:fg<J_2(z)h>:
\end{eqnarray}
These contractions give the following results: 
\begin{equation}
:X^{-}<P^{+}f>gh:\rightarrow -\frac 1{z-w}:(ip^{+}\partial _{p^{+}}f)gh:
\end{equation}
and 
\begin{equation}
\begin{array}{l}
:<P^{+}f>g<X^{-}h>:\rightarrow -\frac{1}{z-w}:fg(ip^{+}\partial _{p^{+}}h):
\\ 
\quad \quad +:fg\left( -ip^{+}p^{-}\int^{w}\frac{dw^{\prime }}{k}\frac{1}{%
z-w^{\prime }}\frac{1}{w-w^{\prime }}:e^{\frac{2u(w^{\prime })}{k}}:\left( 
\frac{w-w^{\prime }}{\sqrt{ww^{\prime }}}\right) ^{\frac{2it_{2}}{k}%
}\right)h:
\end{array}
\end{equation}
and 
\begin{equation}
:f<{P_2}g>h:\rightarrow -\frac 1{z-w}:f((ip^{+}\partial _{p^{+}}+\frac
i2+s)g)h:
\end{equation}
and 
\begin{equation}
\begin{array}{l}
:fg<J_{2}(z)h>:\rightarrow \\ 
\quad \quad :fg\left( ip^{+}p^{-}\int^{w}\frac{dw^{\prime }}{k}\frac{1}{%
z-w^{\prime }}\frac{1}{w-w^{\prime }}:e^{\frac{2u(w^{\prime })}{k}}:\left( 
\frac{w-w^{\prime }}{\sqrt{ww^{\prime }}}\right) ^{\frac{2it_{2}}{k}}\right)
h:
\end{array}
\end{equation}
In the second and the last contractions we have used the relation 
\begin{equation}
\frac 1{(z-w^{\prime })(w-w^{\prime })}=\frac 1{z-w}\left( \frac
1{w-w^{\prime }}-\frac 1{z-w^{\prime }}\right)  \label{rel2}
\end{equation}
Combining these terms we get 
\begin{eqnarray}
J_2(z)\times V_{p^{+}p^{-}}^{r,s}(w) &\rightarrow &-\frac
1{z-w}(ip^{+}\partial _{p^{+}}+\frac i2+s):f(w)g(w)h(w):  \nonumber \\
&=&-\frac 1{z-w}<s,p^{+}|t_2V_{\hat{\jmath}}(w)|s,p^{-}>
\end{eqnarray}
The most involved operator product is the operator product of $J^{-}$ with
the vertex operator. It contains the following contractions: 
\begin{eqnarray}
J^{-}(z)\times V_{p^{+}p^{-}}^{r,s}(w) &\rightarrow &-:(X^{-})^2<P^{+}f>gh:
\\
&&-2:X^{-}<P^{+}f>g<X^{-}h>: \\
&&-:<P^{+}f>g<(X^{-})^2h>: \\
&&-2:X^{-}f<{P_2}g>h:-2:f<{P_2}g><X^{-}h>: \\
&&+:fg<J^{-}(z)h>:
\end{eqnarray}
where in the second contraction the factor of $2$ comes from the
permutations of $X^{-}$'s. The last contraction is equal to 
\begin{eqnarray}
&:&fg<J^{-}(z)h>:\rightarrow  \nonumber \\
&:&fg\left( -p^{-}\int^wdw^{\prime }\frac 1{z-w^{\prime }}:\partial
_{w^{\prime }}e^{\frac{2u(w^{\prime })}k}:\left( \frac{w-w^{\prime }}{\sqrt{%
ww^{\prime }}}\right) ^{\frac{2it_2}k}\right) h:  \label{Jmh1} \\
-2p^{+} &:&(\partial _{p^{+}}f)g\left( -p^{-}\int^w\frac{dw^{\prime }}k\frac
1{z-w^{\prime }}\frac 1{w-w^{\prime }}:e^{\frac{2u(w^{\prime })}k}:\left( 
\frac{w-w^{\prime }}{\sqrt{ww^{\prime }}}\right) ^{\frac{2it_2}k}\right) h:
\label{Jmh2} \\
+ &:&fg\left( -p^{-}\int^wdw^{\prime }\frac 1{(z-w^{\prime })^2}:e^{\frac{%
2u(w^{\prime })}k}:\left( \frac{w-w^{\prime }}{\sqrt{ww^{\prime }}}\right) ^{%
\frac{2it_2}k}\right) h:  \label{Jmh3}
\end{eqnarray}
Partial integration of the third term gives 
\begin{eqnarray}
- &:&fg\left[ p^{-}\frac 1{z-w^{\prime }}:e^{\frac{2u(w^{\prime })}k}:\left( 
\frac{w-w^{\prime }}{\sqrt{ww^{\prime }}}\right) ^{\frac{2it_2}k}\right]
_{w^{\prime }=w}h:  \label{Jmh31} \\
- &:&fg\left( -p^{-}\int^wdw^{\prime }\frac 1{z-w^{\prime }}:\partial
_{w^{\prime }}e^{\frac{2u(w^{\prime })}k}:\left( \frac{w-w^{\prime }}{\sqrt{%
ww^{\prime }}}\right) ^{\frac{2it_2}k}\right) h:  \label{Jmh32} \\
- &:&fg\left( -p^{-}\int^wdw^{\prime }\frac 1{z-w^{\prime }}:e^{\frac{%
2u(w^{\prime })}k}:\partial _{w^{\prime }}\left( \frac{w-w^{\prime }}{\sqrt{%
ww^{\prime }}}\right) ^{\frac{2it_2}k}\right) h:  \label{Jmh33}
\end{eqnarray}
where the term (\ref{Jmh32}) cancels the term in (\ref{Jmh1}). The term (\ref
{Jmh31}) can be written as (writing also the end states explicitly) 
\begin{equation}
\begin{array}{l}
-\frac 1{z-w}f(w)<s,p^{+}|\left( :e^{-i\frac{2u(w)}kt_2}:\right) \left( :e^{%
\frac{2u(w)}k}:\right) p^{-}\left( :e^{-it^{-}X^{+}(w)}:\right) |s,p^{-}>
\end{array}
\label{Jmh311}
\end{equation}
If one inserts $t^{-}$ on the right instead of $p^{-}$ and moves it to the
other side of $:e^{-i\frac{2u(w)}kt_2}:$, one gets 
\begin{equation}
\left( :e^{-i\frac{2u(w)}kt_2}:\right) \left( :e^{\frac{2u(w)}k}:\right)
t^{-}=t^{-}\left( :e^{-i\frac{2u(w)}kt_2}:\right)  \label{t2t-}
\end{equation}
Therefore the contraction of $J^{-}(z)$ with only the last exponential piece
of $V_{p^{+}p^{-}}^{r,s}(w)$ gives us 
\begin{eqnarray}
-\frac 1{z-w} &:&\left( p^{+}f(\partial _{p^{+}}^2g)h+(1-2is)f(\partial
_{p^{+}}g)h+\left( -\frac{kr}{p^{+}}\right) fgh\right) : \\
-2p^{+} &:&(\partial _{p^{+}}f)g\left( -p^{-}\int^w\frac{dw^{\prime }}k\frac
1{z-w^{\prime }}\frac 1{w-w^{\prime }}:e^{\frac{2u(w^{\prime })}k}:\left( 
\frac{w-w^{\prime }}{\sqrt{ww^{\prime }}}\right) ^{\frac{2it_2}k}\right) h:
\\
- &:&fg\left( -p^{-}\int^wdw^{\prime }\frac 1{z-w^{\prime }}:e^{\frac{%
2u(w^{\prime })}k}:\partial _{w^{\prime }}\left( \frac{w-w^{\prime }}{\sqrt{%
ww^{\prime }}}\right) ^{\frac{2it_2}k}\right) h:
\end{eqnarray}
where in the first line the representation of $t^{-}$ on $<s,p^{+}|$ (\ref
{trep}) is used.

The other contractions are as follows 
\begin{equation}
:(X^{-})^2<P^{+}f>gh:\rightarrow \frac 1{z-w}:p^{+}(\partial _{p^{+}}^2f)gh:
\end{equation}
and 
\begin{equation}
\begin{array}{l}
2:X^{-}<P^{+}f>g<X^{-}h>:\rightarrow \frac{1}{z-w}:2p^{+}(\partial
_{p^{+}}f)g(\partial _{p^{+}}h): \\ 
\quad \quad -2p^{+}:(\partial _{p^{+}}f)g\left( -p^{-}\int^{w}\frac{%
dw^{\prime }}{k}\frac{1}{z-w^{\prime }}\frac{1}{w-w^{\prime }}:e^{\frac{%
2u(w^{\prime })}{k}}:\left( \frac{w-w^{\prime }}{\sqrt{ww^{\prime }}}\right)
^{\frac{2it_{2}}{k}}\right) h:
\end{array}
\label{JmV2}
\end{equation}
and 
\begin{equation}
:<P^{+}f>g<(X^{-})^2h>:\rightarrow \frac 1{z-w}:p^{+}fg(\partial
_{p^{+}}^2h):
\end{equation}
and 
\begin{equation}
2:X^{-}f<{P_2}g>h:\rightarrow \frac 1{z-w}:2p^{+}(\partial
_{p^{+}}f)(\partial _{p^{+}}g)h+(1-2is)(\partial _{p^{+}}f)gh:
\end{equation}
and 
\begin{equation}
\begin{array}{l}
2:f<{P_{2}}g><X^{-}h>:\rightarrow \\ 
\quad \quad \frac{1}{z-w}:2p^{+}f(\partial _{p^{+}}g)(\partial
_{p^{+}}h)+(1-2is)fg(\partial _{p^{+}}h): \\ 
\quad \quad +:fg\left( -p^{-}\int^{w}dw^{\prime }\frac{1}{z-w^{\prime }}:e^{%
\frac{2u(w^{\prime })}{k}}:\frac{2it_{2}}{k}\left[ \frac{1}{w-w^{\prime }}+ 
\frac{1}{2z}\right] \left( \frac{w-w^{\prime }}{\sqrt{ww^{\prime }}}\right)
^{\frac{2it_{2}}{k}}\right) h:
\end{array}
\label{JmV5}
\end{equation}
In writing (\ref{JmV2}) and (\ref{JmV5}) we again used the relation (\ref
{rel2}).

Combining these terms we get the desired result 
\begin{eqnarray}
J^{-}(z)\times V_{p^{+}p^{-}}^{r,s}(w) &\rightarrow &-\frac
1{z-w}(p^{+}\partial _{p^{+}}^2+(1-2is)\partial _{p^{+}}-\frac{kr}{p^{+}}%
):fgh: \\
&=&-\frac 1{z-w}<s,p^{+}|t^{-}V_{\hat{\jmath}}(w)|s,p^{-}>
\end{eqnarray}

Next, we calculate the operator product of the vertex operator with the
energy-momentum tensor (\ref{enmom}) and determine its conformal weight. The
expected result of this operator product is 
\begin{equation}
T(z)\times V_{p^{+}p^{-}}^{r,s}(w)\rightarrow \frac{-r+\frac 1k\left( \frac
14+s^2\right) }{(z-w)^2}V_{p^{+}p^{-}}^{r,s}(w)+\frac 1{z-w}\partial
_wV_{p^{+}p^{-}}^{r,s}(w)
\end{equation}
where the right hand side may be written as 
\begin{eqnarray}
\frac{1/k}{(z-w)^2} &<&jp^{+}|\left( (t_2)^2-it_2+t^{+}t^{-}\right) V_{\hat{%
\jmath}}(w)|jp^{-}>  \label{TV1} \\
+\frac 1{z-w} &:&\left( (\partial _wf)gh:+:f(\partial _wg)h:+fg\left(
\partial _wh\right) \right)  \label{TV2}
\end{eqnarray}
and the last term has the form 
\begin{eqnarray}
\frac{fg\left( \partial _wh\right) }{z-w} &=&:\frac{fg}{z-w}\left[ \frac{%
-p^{-}}k\left( P^{+}(w^{\prime })+\frac{p^{+}}{2w^{\prime }}+\frac{p^{+}}{%
w-w^{\prime }}\right) :e^{\frac{2u(w^{\prime })}k}:\left( \frac{w-w^{\prime }%
}{\sqrt{ww^{\prime }}}\right) ^{\frac{2it_2}k}\right] _{w^{\prime }=w}h: 
\nonumber \\
\,  \label{TV3} \\
&&-:\frac{fg}{z-w}\left( -p^{-}\int^w\frac{dw^{\prime }}k\frac{p^{+}}{%
(w-w^{\prime })^2}:e^{\frac{2u(w^{\prime })}k}:\left( \frac{w-w^{\prime }}{%
\sqrt{ww^{\prime }}}\right) ^{\frac{2it_2}k}\right) h:  \label{TV4} \\
&&+:\frac{fg}{z-w}\left( -p^{-}\int^w\frac{dw^{\prime }}k(\cdots ):e^{\frac{%
2u(w^{\prime })}k}:\frac{2it_2/k}{w-w^{\prime }}\left( \frac{w-w^{\prime }}{%
\sqrt{ww^{\prime }}}\right) ^{\frac{2it_2}k}\right) h:  \label{TV5} \\
&&-:\frac{fg}{z-w}\left( -p^{-}\int^w\frac{dw^{\prime }}k(\cdots ):e^{\frac{%
2u(w^{\prime })}k}:\frac{it_2/k}w\left( \frac{w-w^{\prime }}{\sqrt{%
ww^{\prime }}}\right) ^{\frac{2it_2}k}\right) h:  \label{TV6}
\end{eqnarray}
where $(\cdots )$ in the last two terms stand for $(P^{+}(w^{\prime })+\frac{%
p^{+}}{2w^{\prime }}+\frac{p^{+}}{w-w^{\prime }})$.

In this operator product there are the following contractions to be
calculated and then compared to the above result 
\begin{eqnarray}
T(z)\times V_{p^{+}p^{-}}^{r,s}(w) &\rightarrow &:i\partial
_zX^{-}<P^{+}f>gh:+:iP^{+}fg<\partial _zX^{-}h>: \\
&&+:i<P^{+}f>g<\partial _zX^{-}h>: \\
&&+:f<T_{P_2}(z)g>h:+:fg<T_{P_2}(z)h>: \\
&&+\frac 2k :f<{P_2}g><{P_2}h>:
\end{eqnarray}
where the factor of $2$ in the last term comes from the permutations of $P_2$%
's. The result of these contractions are 
\begin{equation}
:i\partial _zX^{-}<P^{+}f>gh:\rightarrow -\frac{ip^{+}}{z-w}:\partial
_wX^{-}fgh:=\frac 1{z-w}:(\partial _wf)gh:  \label{tv1}
\end{equation}
and 
\begin{equation}
:iP^{+}fg<\partial _zX^{-}h>:\rightarrow :fg\left( -p^{-}\int^w\frac{%
dw^{\prime }}k\frac{P^{+}(z)}{(z-w^{\prime })^2}:e^{\frac{2u(w^{\prime })}%
k}:\left( \frac{w-w^{\prime }}{\sqrt{ww^{\prime }}}\right) ^{\frac{2it_2}%
k}\right) h:  \label{tv2}
\end{equation}
and 
\begin{equation}
\begin{array}{l}
:i<P^{+}f>g<\partial _{z}X^{-}h>:\rightarrow \\ 
\quad \quad \quad i\left( \frac{p^{+}}{2z}-\frac{p^{+}}{z-w}\right)
:fg\left( -p^{-}\int^{w}\frac{dw^{\prime }}{k}\frac{-i}{(z-w^{\prime })^{2}}
:e^{\frac{2u(w^{\prime })}{k}}:\left( \frac{w-w^{\prime }}{\sqrt{ww^{\prime
} }}\right) ^{\frac{2it_{2}}{k}}\right) h:
\end{array}
\label{tv3}
\end{equation}
and 
\begin{equation}
:f<T_{P_2}g>h:\rightarrow \frac{1/k}{(z-w)^2}:\left( (t_2)^2-it_2\right)
fgh:+\frac 1{z-w}:f(\partial _wg)h:  \label{tv4}
\end{equation}
and 
\begin{equation}
:fg<T_{P_2}h>:\rightarrow :fg\left( -p^{-}\int^w\frac{dw^{\prime }}k(\cdots
)\frac 1{z-w^{\prime }}:\partial _{w^{\prime }}e^{\frac{2u(w^{\prime })}%
k}:\left( \frac{w-w^{\prime }}{\sqrt{ww^{\prime }}}\right) ^{\frac{2it_2}%
k}\right) h:  \label{tv5}
\end{equation}
and 
\begin{eqnarray}
\frac 2k &:&f<{P_2}g><{P_2}h>:\rightarrow  \nonumber \\
\frac{2t_2}k\left( \frac 1{2z}-\frac 1{z-w}\right) &:&fg\left( -p^{-}\int^w%
\frac{dw^{\prime }}k(\cdots )\left( \frac i{2z}-\frac i{z-w^{\prime
}}\right) :e^{\frac{2u(w^{\prime })}k}:\left( \frac{w-w^{\prime }}{\sqrt{%
ww^{\prime }}}\right) ^{\frac{2it_2}k}\right) h:  \nonumber \\
\,  \label{tv62} \\
=\frac{2it_2}k &:&fg\left( -p^{-}\int^w\frac{dw^{\prime }}k(\cdots )\frac
1{(z-w)(z-w^{\prime })}:e^{\frac{2u(w^{\prime })}k}:\left( \frac{w-w^{\prime
}}{\sqrt{ww^{\prime }}}\right) ^{\frac{2it_2}k}\right) h:  \nonumber \\
\,  \label{tv63} \\
-\frac{2it_2}k &:&fg\left( -p^{-}\int^w\frac{dw^{\prime }}k(\cdots )\frac
1{2z(z-w^{\prime })}:e^{\frac{2u(w^{\prime })}k}:\left( \frac{w-w^{\prime }}{%
\sqrt{ww^{\prime }}}\right) ^{\frac{2it_2}k}\right) h:  \label{tv64} \\
-\frac{2it_2}k &:&fg\left( -p^{-}\int^w\frac{dw^{\prime }}k(\cdots )\frac
1{2z(z-w)}:e^{\frac{2u(w^{\prime })}k}:\left( \frac{w-w^{\prime }}{\sqrt{%
ww^{\prime }}}\right) ^{\frac{2it_2}k}\right) h:  \label{tv65}
\end{eqnarray}
where $(\cdots )$ in the last two contractions stand for $(P^{+}(w^{\prime
})+\frac{p^{+}}{2w^{\prime }}+\frac{p^{+}}{w-w^{\prime }})$.

As one notices the first three terms in $T\times V_{p^{+}p^{-}}^{r,s}$ (\ref
{TV1}-\ref{TV2}), except $\frac{1/k}{(z-w)^2}<t^{+}t^{-}V_{\hat{\jmath}}(w)>$
piece, is given by the contractions (\ref{tv1}) and (\ref{tv4}). The term in
(\ref{tv63}) is the last term in $T\times V_{p^{+}p^{-}}^{r,s}$ (\ref{TV6}).
Adding the contractions (\ref{tv2}) and (\ref{tv3}) one gets 
\begin{eqnarray}
&:&fg\left( -p^{-}\int^w\frac{dw^{\prime }}k\left( P^{+}(z)+\frac{p^{+}}{2z}-%
\frac{p^{+}}{z-w}\right) \frac 1{(z-w^{\prime })^2}:e^{\frac{2u(w^{\prime })}%
k}:\left( \frac{w-w^{\prime }}{\sqrt{ww^{\prime }}}\right) ^{\frac{2it_2}%
k}\right) h:  \label{tv231} \\
=&:&fg\left( -p^{-}\int^w\frac{dw^{\prime }}k\left( P^{+}(z)+\frac{p^{+}}{2z}%
\right) \frac 1{(z-w^{\prime })^2}:e^{\frac{2u(w^{\prime })}k}:\left( \frac{%
w-w^{\prime }}{\sqrt{ww^{\prime }}}\right) ^{\frac{2it_2}k}\right) h:
\label{tv232} \\
- &:&fg\left( -p^{-}\int^w\frac{dw^{\prime }}k\frac{p^{+}}{z-w}\frac
1{(w-w^{\prime })^2}:e^{\frac{2u(w^{\prime })}k}:\left( \frac{w-w^{\prime }}{%
\sqrt{ww^{\prime }}}\right) ^{\frac{2it_2}k}\right) h:  \label{tv233} \\
+ &:&fg\left( -p^{-}\int^w\frac{dw^{\prime }}k\frac{p^{+}}{(z-w^{\prime
})(w-w^{\prime })}\left( \frac 1{z-w^{\prime }}+\frac 1{w-w^{\prime
}}\right) :e^{\frac{2u(w^{\prime })}k}:\left( \frac{w-w^{\prime }}{\sqrt{%
ww^{\prime }}}\right) ^{\frac{2it_2}k}\right) h:  \nonumber \\
\,  \label{tv234}
\end{eqnarray}
The second term in the above expression is the term in $T\times
V_{p^{+}p^{-}}^{r,s}$ (\ref{TV4}).

The term in (\ref{tv63}) can be written as 
\begin{eqnarray}
&:&\frac{2it_2fg}{k\left( z-w\right) }\left( -p^{-}\int^w\frac{dw^{\prime }}%
k(\cdots )\frac 1{w-w^{\prime }}:e^{\frac{2u(w^{\prime })}k}:\left( \frac{%
w-w^{\prime }}{\sqrt{ww^{\prime }}}\right) ^{\frac{2it_2}k}\right) h:
\label{tv611} \\
+ &:&fg\left( -p^{-}\int^w\frac{dw^{\prime }}k(\cdots )\frac 1{z-w^{\prime
}}:e^{\frac{2u(w^{\prime })}k}:\left( \frac 1{\sqrt{ww^{\prime }}}\right) ^{%
\frac{2it_2}k}\partial _{w^{\prime }}(w-w^{\prime })^{\frac{2it_2}k}\right)
h:  \label{tv612}
\end{eqnarray}
by using the relation (\ref{relation}). The term in (\ref{tv611}) gives the
term in (\ref{TV5}). Partial integration of the second term above gives 
\begin{equation}
\begin{array}{l}
\quad :fg\left[ \frac{-p^{-}}k\left( P^{+}(w^{\prime })+\frac{p^{+}}{%
2w^{\prime }}+\frac{p^{+}}{w-w^{\prime }}\right) \frac 1{z-w^{\prime }}:e^{%
\frac{2u(w^{\prime })}k}:\left( \frac{w-w^{\prime }}{\sqrt{ww^{\prime }}}%
\right) ^{\frac{2it_2}k}\right] _{w^{\prime }=w}h: \\ 
+:fg\left( -p^{-}\int^w\frac{dw^{\prime }}k\left( P^{+}(w^{\prime })+\frac{%
p^{+}}{2w^{\prime }}+\frac{p^{+}}{w-w^{\prime }}\right) \frac 1{z-w^{\prime
}}:e^{\frac{2u(w^{\prime })}k}:\frac{it_2/k}{w^{\prime }}\left( \frac{%
w-w^{\prime }}{\sqrt{ww^{\prime }}}\right) ^{\frac{2it_2}k}\right) h: \\ 
-:fg\left( -p^{-}\int^w\frac{dw^{\prime }}k\left( P^{+}(w^{\prime })+\frac{%
p^{+}}{2w^{\prime }}+\frac{p^{+}}{w-w^{\prime }}\right) \frac 1{(z-w^{\prime
})^2}:e^{\frac{2u(w^{\prime })}k}:\left( \frac{w-w^{\prime }}{\sqrt{%
ww^{\prime }}}\right) ^{\frac{2it_2}k}\right) h: \\ 
-:fg\left( -p^{-}\int^w\frac{dw^{\prime }}k\left( \partial _{w^{\prime
}}P^{+}(w^{\prime })-\frac{p^{+}}{2(w^{\prime })^2}+\frac{p^{+}}{%
(w-w^{\prime })^2}\right) \frac 1{z-w^{\prime }}:e^{\frac{2u(w^{\prime })}%
k}:\left( \frac{w-w^{\prime }}{\sqrt{ww^{\prime }}}\right) ^{\frac{2it_2}%
k}\right) h: \\ 
-:fg\left( -p^{-}\int^w\frac{dw^{\prime }}k\left( P^{+}(w^{\prime })+\frac{%
p^{+}}{2w^{\prime }}+\frac{p^{+}}{w-w^{\prime }}\right) \frac 1{z-w^{\prime
}}:\partial _{w^{\prime }}e^{\frac{2u(w^{\prime })}k}:\left( \frac{%
w-w^{\prime }}{\sqrt{ww^{\prime }}}\right) ^{\frac{2it_2}k}\right) h:
\end{array}
\label{tv632}
\end{equation}
The second term above cancels the term in (\ref{tv64}) and the last term
cancels (\ref{tv5}). Combining the third and the fourth terms with the terms
of (\ref{tv232},\ref{tv234}) one gets finite terms. Therefore, after all the
calculations done so far we obtained all the terms in $T\times
V_{p^{+}p^{-}}^{r,s}$ (\ref{TV1}-\ref{TV6}), except the $\frac{1/k}{(z-w)^2}%
<t^{+}t^{-}V_{\hat{\jmath}}(w)>$ piece and the term (\ref{TV3}) and the only
remaining term coming from the contractions is the first term in (\ref{tv632}%
). This term can be written as 
\begin{equation}
\begin{array}{ll}
& \frac 1{z-w}:fg\left[ \frac{-p^{-}}k\left( P^{+}(w^{\prime })+\frac{p^{+}}{%
2w^{\prime }}\right) :e^{\frac{2u(w^{\prime })}k}:\left( \frac{w-w^{\prime }%
}{\sqrt{ww^{\prime }}}\right) ^{\frac{2it_2}k}\right] _{w^{\prime }=w}h: \\ 
+ & \frac 1{z-w}:fg\left[ \frac{-p^{-}}k\frac{p^{+}}{w-w^{\prime }}:e^{\frac{%
2u(w^{\prime })}k}:\left( \frac{w-w^{\prime }}{\sqrt{ww^{\prime }}}\right) ^{%
\frac{2it_2}k}\right] _{w^{\prime }=w}h: \\ 
- & \frac 1{z-w}:fg\left[ \frac{-p^{-}}k\frac{p^{+}}{z-w^{\prime }}:e^{\frac{%
2u(w^{\prime })}k}:\left( \frac{w-w^{\prime }}{\sqrt{ww^{\prime }}}\right) ^{%
\frac{2it_2}k}\right] _{w^{\prime }=w}h:
\end{array}
\label{tv6321}
\end{equation}
In writing this expression we used the relation (\ref{rel2}). The
combination of first two terms in (\ref{tv6321}) gives the term (\ref{TV3}).
Whereas the last term in (\ref{tv6321}) can be written as 
\begin{equation}
\frac{1/k}{(z-w)^2}<s,p^{+}|p^{+}f(w)\left( :e^{-i\frac{2u(w)}kt_2}:\right)
\left( :e^{\frac{2u(w)}k}:\right) \left( :e^{-it^{-}X^{+}}:\right)
p^{-}|s,p^{-}>
\end{equation}
Changing $p^{+}$ to $t^{+}$ and $p^{-}$ to $t^{-}$, and then moving $t^{-}$
past $t_2$ (\ref{t2t-}) one gets 
\begin{equation}
\frac 1{(z-w)^2}<s,p^{+}|\frac{t^{+}t^{-}}kV^{r,s}(w)|s,p^{-}>
\end{equation}
This is the last piece needed in the operator product of the vertex operator
with the energy-momentum tensor (\ref{TV1}-\ref{TV6}).

This computation proves that we have constructed the correct vertex operator
at the quantum level.

\end{document}